\documentclass[useAMS,usenatbib]{mn2e}
\usepackage{pdfsync}
\usepackage{amsmath,amssymb}
\usepackage{bm}

\usepackage{delarray}
\usepackage{graphicx}

\def\psfig#1{{}}        

\newcommand{\mathd}{\mathrm{d}}

\newcommand{\Tmin}{{t_\mathrm{min}}}
\newcommand{\Tmax}{{t_\mathrm{max}}}

\newcommand{\vmin}{{\v_\mathrm{min}}}
\newcommand{\vmax}{{\v_\mathrm{max}}}
\newcommand{\umin}{{u_\mathrm{min}}}
\newcommand{\umax}{{u_\mathrm{max}}}
\newcommand{\fo}{{\cal{F}}}
\newcommand{\diag}{\mathrm{diag}} 

\newcommand{\Frest}{F_\mathrm{rest}} 

\newcommand{\fext}{f_\mathrm{ext}}
\newcommand{\Fext}{\mathbf{f}_\mathrm{ext}}

\def\v {{\rm{v}}}
\def\V {{\Xi}}

\newcommand{\Msun}{\ensuremath{\mathrm{M}_{\odot}}}

\def\Eq#1{{Eq.~(\ref{e:#1})}}   

\def\Tab#1{{Table~\ref{t:#1}}}        
\def\Fig#1{{Fig.~\ref{f:#1}}}

\newcommand{\Sec}[1]{Sect.~\ref{s:#1}}
\def\be{\begin{equation}}
\def\ee{\end{equation}}
\def\ba{\begin{eqnarray}}
\def\ea{\end{eqnarray}}

\def\pier#1{\noindent{\bf[$\clubsuit$ #1]}}

\def\pdrv#1#2{\frac{\partial #1}{\partial #2}}  

\def\M#1{{\mathbf{#1}}} 
\def\T#1{{{#1}^{\top}}}         
\def\MG#1{\bm{#1}} 

\def\la{\lambda}

\def\for{{\cal{F}}^\ast}

\def\AMR{{age-metallicity relation}}
\def\SSP{{single stellar population}}
\def\SSPs{{single stellar populations}}
\def\SAD{{stellar age distribution}}

\def\SED{{spectral energy distribution}}

\def\PHR{{\sc P\'egase-HR}}

\def\GCV{{generalized cross validation}}

\newcommand{\nicefrac}[2]{\leavevmode\kern.1em
            \raise.5ex\hbox{\the\scriptfont0 #1}\kern-.1em
      /\kern-.15em\lower.25ex\hbox{\the\scriptfont0 #2}}

\title{STEllar Content and Kinematics from high resolution galactic spectra via Maximum A Posteriori}

\author[P. Ocvirk,  C. Pichon, A. Lan\c{c}on, E. Thi\'ebaut]{  P. Ocvirk$^{{1}}$, C. Pichon$^{{2}}$, A. Lan\c{c}on$^{{1}}$, \& E.~Thi\'ebaut$^{{3}}$\\
$^1$Observatoire de Strasbourg (UMR 7550), 11 rue de l'Universit\'e,
67000 Strasbourg, France. \\
$^2$Institut d'Astrophysique de Paris, 98 bis boulevard Arago, 75014 Paris, France. UMR 7095\\
$^3$Observatoire de Lyon, 9 avenue Charles Andr\'e F-69561 Saint Genis Laval
 Cedex, France
}

\begin{document}

\date{Typeset \today ; Received / Accepted}  
\pagerange{\pageref{firstpage}--\pageref{lastpage}} \pubyear{2005}

\maketitle
\label{firstpage}
\begin{abstract}   
We introduce STECKMAP (STEllar Content and {\bf K}inematics via Maximum A Posteriori), a method to recover the kinematical properties  of a galaxy simultaneously
with its stellar content from integrated light spectra. It is an extension of STECMAP \citep{ocvirk05} to the general case
where the velocity distribution of the underlying stars is also unknown.
The reconstructions of the stellar age distribution, the
age-metallicity relation, and the  Line-Of-Sight Velocity Distribution  (LOSVD) are all
non-parametric, {\em i.e.} no specific shape is assumed. The only a
propri we use are positivity and the requirement that the solution is smooth
enough. The smoothness
parameter can be set by {\GCV } according to the level of noise in the data in order to avoid overinterpretation. 

We use single stellar populations (SSP) from {\PHR} ($R=10\,000$,
$\lambda\lambda = 4\,000-6\,800$ {\AA}, \citealt{PEG-HR}) to test the method through
realistic simulations. Non-Gaussianities in LOSVDs are reliably recovered with
SNR as low as $20$ per $0.2$ {\AA } pixel. It turns out that the recovery of the stellar content is not degraded by the simultaneous recovery of the kinematic
distribution, so that the resolution in age and error estimates given in \citet{ocvirk05} remain appropriate when used with STECKMAP.

We 
also  explore  the
case of age-dependent kinematics ({\em i.e.} when each stellar component has its
own LOSVD). We separate the bulge and disk components of an idealized
simplified spiral galaxy in integrated light from high quality pseudo data (SNR=100 per pixel, R=$10\,000$), and constrain the kinematics (mean projected velocity, projected velocity dispersion) and age of both components.

\end{abstract}


%





\section{Introduction}
\label{s:intro}

For decades now, the spectral indices from the Lick group have been used to study the properties of stellar
populations \citep{faber85,worthey94,trager98}. Since the profile and depth of the lines involved in these
spectral indices are affected by the Line Of Sight Velocity Distribution
(hereafter LOSVD) of the stars, it is necessary to correct the measured depths by a
factor depending on the moments of the velocity distribution \citep{davies93,kuntschner2000,kuntschner04}. 
The latter moments must be determined using specialized code
\citep{bender90,saha94,pinkney-bender03,merritt97,kuijken-merrifield93,vdmarel-franx93}. These kinematical deconvolution routines have been
used for some time and have undergone 2 major mutations. First, thanks
to the increasing power of computers, it became affordable to swap back and forth
from direct
space to Fourier space, so that many disturbances such as
 border effects and saturation could be avoided. 
It became
straightforward to mask problematic regions of the data, such as dead pixels,
emission lines, etc... The second evolution of these codes allowed the use of multiple superimposed stellar templates
to best match the observed spectrum \citep{rix-white, cappellari04}. It has
also been proposed to use {\SSP s} as synthetic templates, and this approach has
proved to be useful in addressing the template mismatch problem
\citep{falcon-peletier}. Moreover, this technique can
save precious telescope time since it circumvents the need for observing 
template stars.

In \citealt{ocvirk05} (hereafter Paper I), we introduced STECMAP, a method for recovering
non-parametrically the stellar content of a given galaxy from its integrated light spectrum.
Using STECMAP requires, as a preliminary, convolving the
data or models with the proper Point Spread Function (PSF), which can be of both
physical ({\em{i.e.}} the stellar LOSVD) and instrumental (the instrument's PSF)
origin. Adjusting the LOSVD to fit the data does not
only constrain the kinematics of the observed galaxy but will also reduce the
mismatch due to errors in the determination of the redshift or anomalous
PSF, which is ultimately a necessary step when fitting galaxy spectra.
 
Here we propose to constrain the velocity distribution simultaneously with the stellar content, by merging the kinematic deconvolution and the stellar
content reconstruction in one global Maximum A
Posteriori likelihood inversion method. Hence, STECMAP becomes STECKMAP
(STEllar Content and Kinematics via Maximum A Posteriori likelihood). In this
respect, STECKMAP resembles the method proposed by {\it e.g.} \citet{falcon-peletier},
except that it takes advantage of the treatment of the stellar content by STECMAP.
Together with the stellar age distribution and the age-metallicity relation, the LOSVD is described  non-parametrically and the only a priori we use
are smoothness and positivity.

We also tentatively address the case of age-dependent kinematics, {\em i.e.} we
try to  recover the individual LOSVDs and ages of several superimposed kinematical
subcomponents. This approach is motivated by the fact that galaxies often display several
kinematical components. Ellipticals and dwarf ellipticals are for instance
 known to
often harbor kinematically decoupled cores
\citep{des-kdc,balcells90,bender92}, and spiral galaxies are usually assumed to be constituted of a
thin and a thick disk, a bulge and a halo \citep{newgal}. 
The variety of the dynamical properties of the components has a counterpart in their
stellar content, as a signature of the formation and evolution of the galaxy.
For instance, the halo of the Milky Way is
believed to consist mainly of old, metal poor stars, while the bulge is more
metal rich, and the thin disk is
mainly younger than the bulge \citep{newgal}.
It is thus natural to let any stellar sub
population have its own LOSVD. This possibility has been recently addressed by
\citet{debruyne04-1,debruyne04-2}, in a slightly different framework:
they use individual stars as templates for the different components, while we propose to use synthetic SSP
models. Such a method would allow us to separate the several kinematical components of
galaxies from
integrated light spectra, and constrain for example their age-velocity dispersion and
{\AMR}. The highly detailed stellar content and kinematical information that
can be obtained for the Milky Way or for nearby galaxies that can be resolved into stars, such as
M31 \citep{ferguson02,ibata04}, could be extended to a larger sample of more
distant galaxies. This technique could also be useful in detecting and
characterizing major stellar streams in age and velocity from integral field
spectroscopy of galaxies.

In the whole paper we use the {\PHR} SSP models \citep{PEG-HR} in order to illustrate
and investigate the behaviour of the problems through simulations and
inversions of mock data. Indeed, {\PHR}, with its high spectral resolution
($R=10\,000$), is an adequate choice for testing the recovery of detailed
kinematical information in the form of non-parametric LOSVDs. The problems and
methods we describe are however by no means specific to {\PHR} (and its wavelength coverage) and STECKMAP
could be used with any possible SSP model, depending on the type of data that
is being analyzed.

We will start with the modeling of the
kinematics. Then, we will address the idealized linear problem
of recovering the LOSVD when the stellar
content is known, {\em i.e.} the template is assumed to be perfect.
Section \ref{s:azek} deals with simultaneous age and LOSVD reconstruction of composite populations.
Finally, section \ref{s:adk} investigates the case of age-dependent
kinematics in a simplified context where the metallicity and extinction are
known a priori.

\section{Models of galaxy spectra}
\label{s:model}

In this section we present the modeling of galaxy spectra, taking into
account the composite nature of the stellar population, in age, metallicity
and extinction, and finally its kinematics.

\subsection{The composite reddened population at rest}

We model the SED of the composite reddened population at rest using
the ingredients defined in Paper I:
\begin{equation}
\Frest(\lambda) = \fext(E,\lambda) \int_{\Tmin}^{\Tmax}\Lambda(t)
B(\lambda,t,Z(t))\mathd t \, ,
\label{e:evergely}
\end{equation}
where $\Lambda(t)$ is the luminosity weighted stellar age distribution, $Z(t)$
is the age-metallicity relation, and $B(\lambda,t,Z)$ is the flux-averaged
{\SSP } basis of an isochrone population of age $t$, 
$\fext$ the extinction law, and metallicity $Z$. We
recall briefly the main properties of the {\PHR} SSP basis we used in this
paper. As mentioned earlier, spectral resolution is $R=10\,000$ over the full optical domain $\lambda
\lambda =[4000, 6800]$ {\AA}, sampled in steps of $0.2$ {\AA}. The models are available for metallicities $Z
\in [0.0001, \,  0.1]$ and considered reliable between $\Tmin=10$ Myr and
$\Tmax=15$ Gyr.  The IMF used is described in
\cite{kimf} and the stellar masses range from $0.1$\,\Msun\ to
$120$\,\Msun. The extinction law  $\fext$ was taken from \citet{calzetti01}.

\subsection{Model kinematics}
Stellar motions in galaxies 
define a LOSVD corresponding to projected local velocity distributions
integrated along the line of sight and across one resolved spatial element. 

\subsubsection{Global kinematics}
\label{s:sk}

The motion of the stars can to first approximation be accounted for by
assuming that the velocities of all
stars of all ages along the line of sight are taken from the same velocity distribution (hence ``global''). The model {\SED}, ${\phi(\lambda)}$,
is the convolution of the assumed normalized LOSVD, ${g(\v)}$, defined for $\v
\in [\vmin, \vmax]$ with the model spectrum at
rest ${\Frest(\lambda)}$.  The convolved spectrum ${\phi(\la)}$ reads:
\begin{equation}
\phi(\lambda)=\int_{\vmin}^{\vmax} \Frest \left(
\frac{\lambda}{1+\v / c} \right)  g(\v) \frac{\mathd
\v}{{1+ \v / c}} \, ,
\label{e:sk}
\end{equation}
where $c$ is the velocity of light. The above expression reads as a standard convolution 
\begin{equation}
\widetilde{\phi}(w)=c \int_{\umin}^{\umax} \widetilde{F}(w-u) \widetilde{g}(u) \mathd u \, ,
\label{e:rsk}
\end{equation}
with the following reparameterization:
\begin{eqnarray}
w &\equiv& \ln(\lambda) \,, \quad
u \equiv \ln({1+\frac{\v}{c}}) \, , \\
\widetilde{F}(w) &\equiv& \Frest({\rm{e}}^{w})=\Frest(\lambda) \, ,\\
\widetilde{g}(u) &\equiv&  g(c\,({\rm{e}}^{u}-1))=g(\v) \, , \,\,\,\,
\widetilde{\phi}(w) \equiv \phi({\rm{e}}^w)= \phi(\lambda)\, , \\ 
\umin&=&\ln(1+\frac{\vmin}{c}) \, , \,\,\,
\umax=\ln(1+\frac{\vmax}{c}) \, .
\end{eqnarray}

\subsubsection{Age-dependent kinematics}

We now allow the LOSVD to depend on the age of the stars. 
For simplicity, we consider here only unreddened mono-metallic populations,
{\em i.e.} $\fext(E,\la)=1$ and $Z(t)=Z_0$.
We introduce the age-velocity distribution, $\V(\v,t)$, defined in $[\vmin,\vmax] \times [\Tmin,
\Tmax]$, which gives the contribution of stars of velocity and age in
$[\v,\v+\mathd \v] \times [t,t+\mathd t]$ to the total observed light. Thus, for a given
age $t$, $\V(\v,t)$ is the LOSVD of the {\SSP} of age $t$. The age-velocity
distribution $\V(\v,t)$ is related to the stellar age distribution, ${\Lambda(t)}$, by:
\begin{equation}
\int_{\vmin}^{\vmax}\V(\v,t) \mathd \v=\Lambda(t) \, ,
\end{equation}
The model spectrum of such a population thus reads:
\begin{eqnarray}
\phi(\lambda) = \int_{\Tmin}^{\Tmax} \int_{\vmin}^{\vmax} B \left(
\frac{\lambda}{1+\v / c},t,Z_0   \right)  \V(\v,t) \frac{\mathd
\v \, \mathd t}{{1+\v / c}}  ,
\label{e:ck1} 
\end{eqnarray}
The above expression can be rewritten more conveniently
\begin{equation}
\widetilde{\phi}(w)=c \int_{\umin}^{\umax} \int_{\Tmin}^{\Tmax} \widetilde{B}(w-u,t)  \widetilde{\V}(u,t)
\, \mathd t \, \mathd u \, ,
\label{e:ckrs}
\end{equation}
using the same reparameterization as in \Sec{sk} and 
\begin{eqnarray}
\widetilde{B}(w,t)&\equiv& B({\rm{e}}^{w},t,Z_0)=B(\lambda,t,Z_0) \, , \\
\widetilde{\V}(u,t)&\equiv& \V(c\,({\rm{e}}^{u}-1),t)=\V(\v,t) \, .
\end{eqnarray}
%

In the rest of the paper, we will use exclusively the standard ({\em{i.e.}}
reparameterized) convolutions as in \Eq{rsk} and \Eq{ckrs}. For readability, we
will drop the superscript $\widetilde{\,\,\,}$ and set the speed of light to one.

\section{Kinematical deconvolution}
\label{s:kin}
\label{s:npkd}
\Sec{sk} shows that with proper reparameterization, the convolution of a
model spectrum at rest, $F(w)$, with the stellar LOSVD, $g(u)$, reads as a standard convolution, given by \Eq{rsk}.
Finding the LOSVD when the observed spectrum, $\phi(w)$, and the template
spectrum, $F(w)$, are given is a 
classical deconvolution problem. 
Our goal here is not to discuss the respective qualities of the many different
methods available in the literature to solve this problem. Most rely on
fitting the data while imposing some a priori on the LOSVD, {\em i.e.} they
provide Maximum A Posteriori (MAP) estimates of the LOSVD. Let us 
 describe briefly our method
to obtain such a solution with the purpose of coupling it in a later step with
STECMAP.

\subsection{The convolution kernel}

Here we discretize \Eq{rsk} to obtain a matrix form defining
the convolution kernel. We use an evenly spaced set
\begin{displaymath}
  \bigl\{u_{j} = \umin + (j - 1/2)\,\delta{}u;
	j=1,2,\ldots,p\bigr\}
\end{displaymath}
spanning   $[\umin,\umax]$   with   constant  step   $\delta
u\equiv(\umax-\umin)/p$. We expand the LOSVD as a sum of $p$
gate functions:
\begin{displaymath}
  g(u) = \frac{1}{\delta{}u} \,
	\sum_{j}g_{j}\,\theta\left(\frac{u - u_{j}}{\delta{}u}\right)
\end{displaymath}
where
\begin{displaymath}
  \theta(x) = \left\{\begin{array}{ll}
  1 & \mbox{if $-1/2<x\le1/2$}\\
  0 & \mbox{otherwise}\\
  \end{array}\right.\,.
\end{displaymath}
Injecting this expansion into \Eq{rsk} leads to:   
\begin{eqnarray}
	\phi(w) &=&
	  \frac{1}{\delta{}u} \, \sum_{j=1}^{j=p} g_j \,
	  \int\limits_{\umin}^{\umax} F(w - u) \,
		\theta\left(\frac{u - u_{j}}{\delta\,u}\right) \mathd u \, , \nonumber \\  
	&\simeq&
		\sum_{j=1}^{j=p} g_{j} \, F(w - u_{j}).
\end{eqnarray}
Similarly,   we  now   sample  along   the   wavelengths  by
integrating over a small $\delta{}w$:
\begin{eqnarray}
  \phi_{i} &\equiv&
    \frac{1}{\delta{}w} \, \int \phi(w) \,
		\theta\left(\frac{w - w_{i}}{\delta\,w}\right) \mathd w \, , \nonumber \\
	&\simeq&
		\sum_{j=1}^{j=p} g_{j} \, F(w_i - u_{j}) \, ,
		\label{e:skd}
\end{eqnarray}
where    $\{w_{j};   j=1,2,\ldots,m\}$    is   a    set   of
\emph{logarithmic}  wavelengths spanning the  spectral range
with a constant step.

Using matrix notation and accounting for data noise, the observed SED reads:
\begin{equation}
{\M{y}} =\M{K \cdot g} + \M{e} \, ,
\label{e:mk}
\end{equation}
where $\M{y}=\T{(\phi_{1}, \phi_{2}, \dots,\phi_{m})}$ is the measured
spectrum, and ${\M{e}}=\T{(e_1, e_2, \dots, e_m)}$ accounts for modelling
errors and noise. The vector
of sought parameters $\M{g}=\T{(g_{1}, g_{2}, \dots, g_{p})}$ is the
discretized LOSVD. The vector $\M{s=K\cdot g}$ is the {\em model} of the
observed spectrum and the matrix $\M{K}$ 
\begin{equation}
K_{ij}=F(w_{i}-u_{j}) \, ,\,\, \forall
(i,j) \in \{1,\dots,m \} \times \{1, \dots, p \} \, ,
\label{e:defK}
\end{equation}
is called the convolution kernel.

\label{s:fourier}

The convolution theorem
\citep{NR} states that the Fourier transform of the convolution of two functions is equal to the
frequency-wise  product of the individual Fourier transforms of the two functions.
Applying this theorem yields another equivalent expression for the model spectrum $\M{s}$:
\begin{equation}
\M{s}={\fo}^{-1} \cdot \diag (\fo \cdot \M{F}) \cdot \fo \cdot \M{g}\, ,
\label{e:conf}
\end{equation}
where $\fo$ is the discrete Fourier operator defined in \citet{NR} as:
\begin{eqnarray}
\!\! {\cal{F}}_{ij} \!\!  &=& \!\!\!  {\exp}\left({\frac{2 {\rm{i}} \pi}{m} (i-1)(j-1)} \right) , \,
\forall \, (i,j) \in [1,\dots\,m]^2 \, , \\
 {\cal{F}}^{-1} \!\! &=&\!\!\!  \frac{1}{m}  {\cal{F}}^{\ast}  \, .
\label{e:fon}
\end{eqnarray}
Note that since $m$ is the size of the template spectrum $\M{F}$, the
discretized LOSVD
$\M{g}$, which is initially of size $p$ needs to be symmetrically padded with
zeros to the size $m$  in order to transform the Toeplitz matrix into a circulant
one.
The diagonal matrix $\diag(\fo \cdot \M{F})$ carries the coefficients of the Fourier
transform of the model spectrum at rest $\M{F}$. This notation involving the
Fourier operator, ${\cal{F}}$, will be very
useful for a number of algebraic derivations in the rest of the
paper. In practice, from a computational point of view, it is more efficient to implement any forward or
inverse Fourier transform through FFT. Similarly, the product $\diag (\fo
\cdot \M{F}) \cdot \fo \cdot \M{g}$  is  in practice implemented as a
frequency-wise product of the individual FFTs.

\subsection{Regularization and MAP}

A number of earlier publications have shown that the maximum likelihood solution to \Eq{mk}
is  very sensitive to the noise in the data $\M{e}$. 
Hence, in the spirit of Paper I,
 we choose to regularize the problem by requiring the LOSVD to be smooth. To do so, we
use the quadratic penalization $P(\M{g})$ as defined by
Eq.~(29) in Paper I:
\begin{equation}
P(\M{g})=\T{\M{g}} \cdot \T{\M{L}} \cdot \M{L \cdot g} \, .
\label{e:penal}
\end{equation}
In the rest of
the paper, the penalization is Laplacian, {\em i.e.} $\M{L}=\M{D}_2$, 
where $\M{D}_{2}$ is the discrete second order difference operator, as defined
in \citet{pichsieb02}. The objective function, $Q_{\mu}$,
 to be minimized is given by:
\begin{equation}
Q_{\mu}(\M{g})=\chi^2(\M{y|g}) + \mu P(\M{g}) \, ,
\label{e:defQ}
\end{equation}
where the $\chi^2$ is defined by 
\begin{equation}
  \chi^2(\M{y}|\M{g}) =
    \T{\bigl(\M{y} - \M{s}(\M{g})\bigr)}{\cdot}\M{W}{\cdot} 
       \bigl(\M{y} - \M{s}(\M{g})\bigr) \, .
   \label{e:chi2}
\end{equation}
The vector $\M{y}$ is the observed spectrum and the weight matrix is the inverse of the covariance matrix of the
noise: $\M{W}=\mathrm{Cov}(\M{e})^{-1}$. The parameter $\mu$ controls the smoothness of the LOSVD through its coefficients, $\M{g}$. 
It can be set on the basis of
simulations (as described in Paper I) or automatically by GCV \citep{wahba}, according to the SNR of the data. In the latter case, the properties of the convolution kernel
can be used to speed up the computation of the GCV function.
Further regularization is provided by the requirement of positivity, implemented
through quadratic reparameterization. Minimizing $Q_\mu$ yields the
regularized solution $\M{g}_\mu$. Efficient minimization procedures require the
analytical expression of the gradients of $Q_\mu$, given in \Sec{gradkd}.

\subsection{Simulations}

\label{s:sims}

\begin{figure}
{\includegraphics[width=0.96\linewidth,clip]{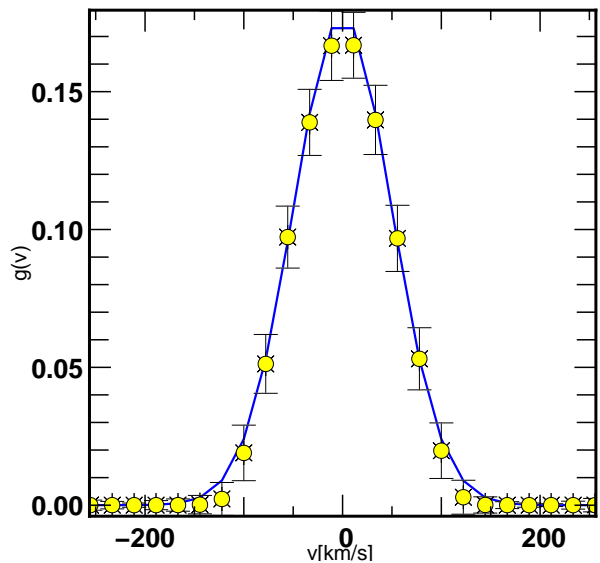}}
\caption{Non parametric reconstruction of a Gaussian LOSVD for simulated data,
$\sigma_{v}=100$
km/s, SNR=10 per pixel. The
model spectrum at rest is a $10$ Gyr old solar metallicity {\SSP} with $R=10\,000$ at
$4000 - 6800$ \AA. The template spectrum is identical, so that no template
mismatch is allowed here. The thick line is the 
input model. The circles and the bars show respectively the median and the
 interquartiles of the recovered solutions for 500 realizations of the noise.  
}
\label{f:lossim}
\end{figure}

\begin{figure*}
\begin{center}
\begin{tabular}{ll}
\includegraphics[width=0.4\linewidth,clip]{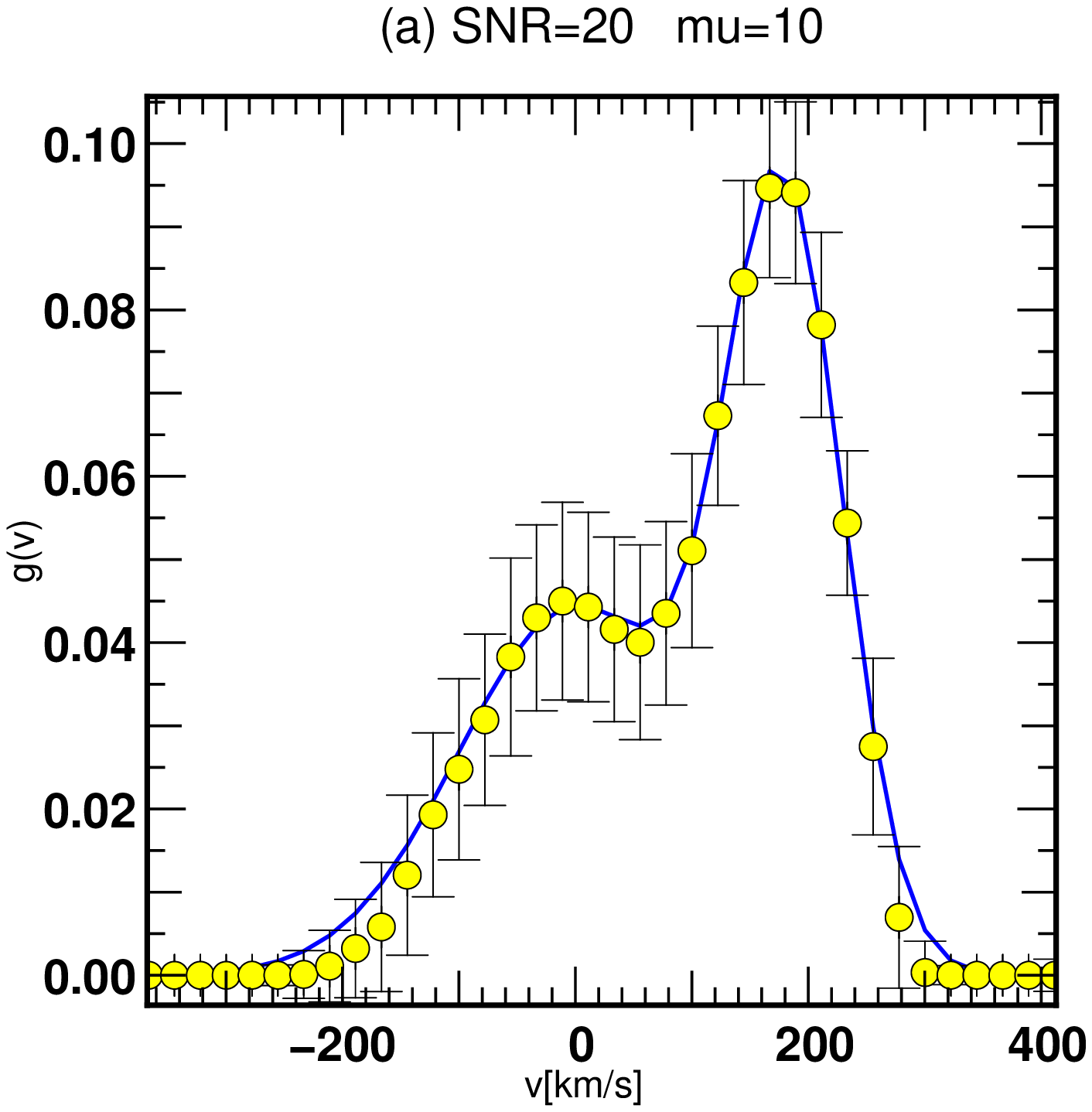} &
\includegraphics[width=0.4\linewidth,clip]{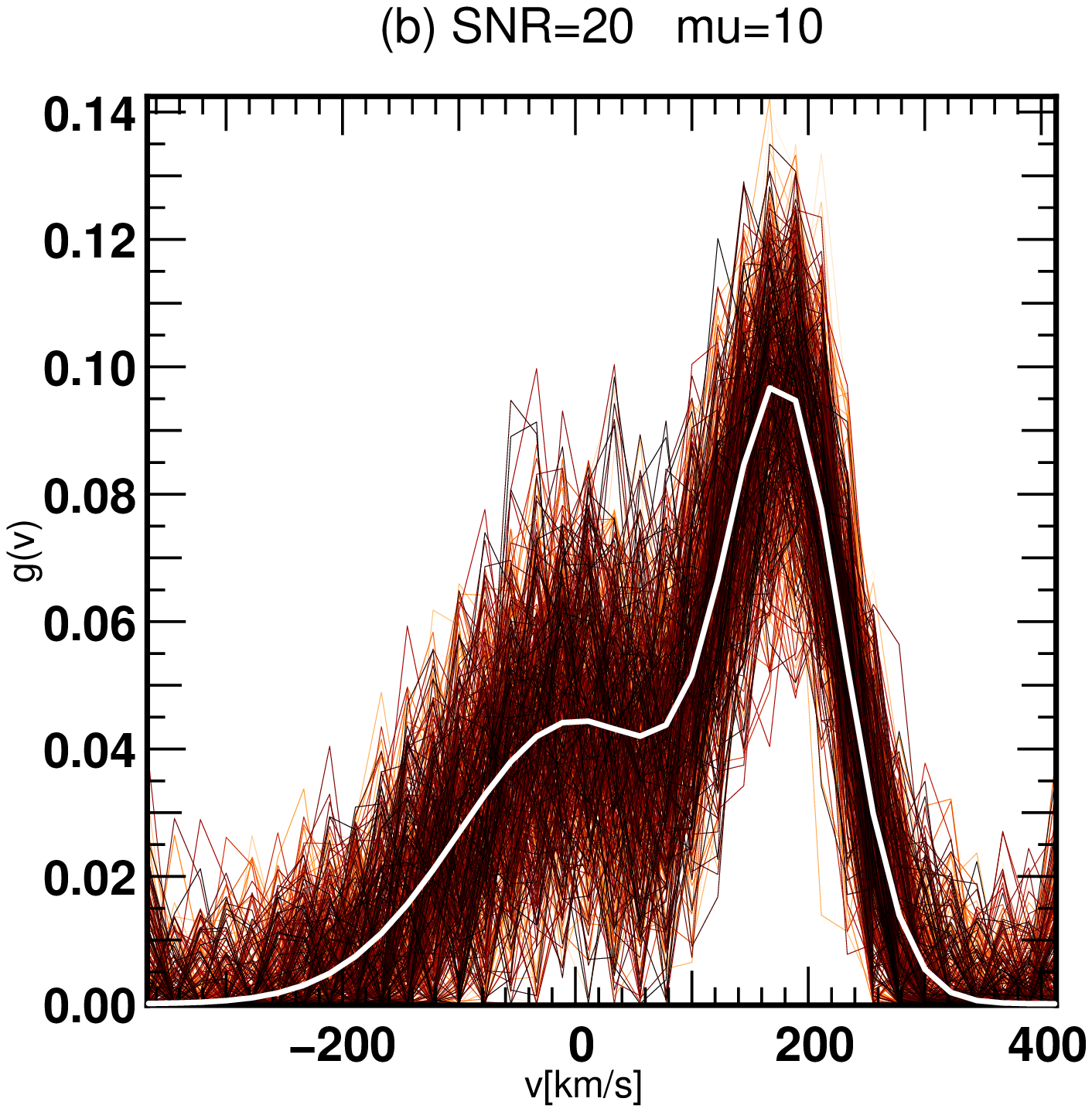} \\
\includegraphics[width=0.4\linewidth,clip]{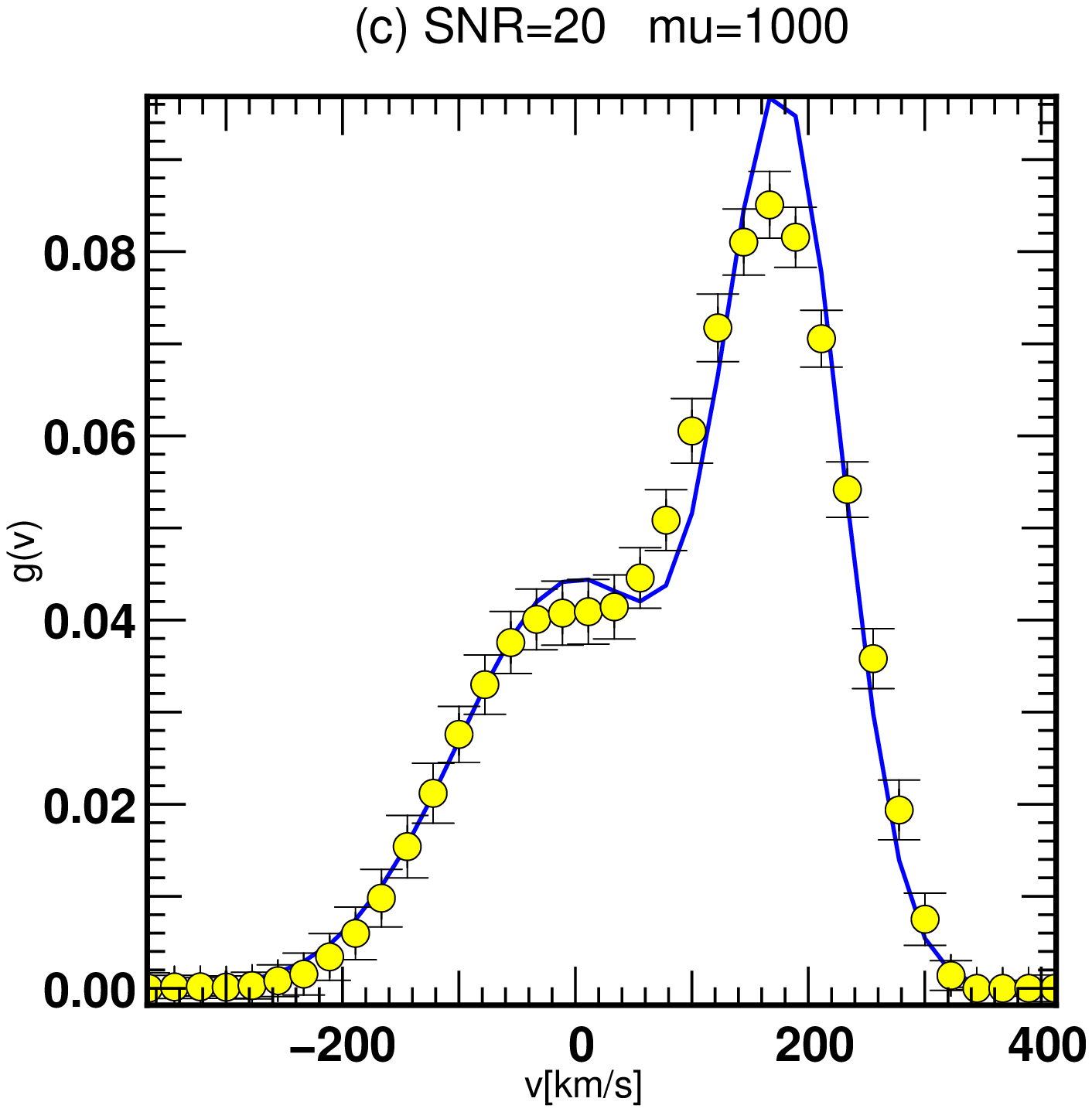} &
\includegraphics[width=0.4\linewidth,clip]{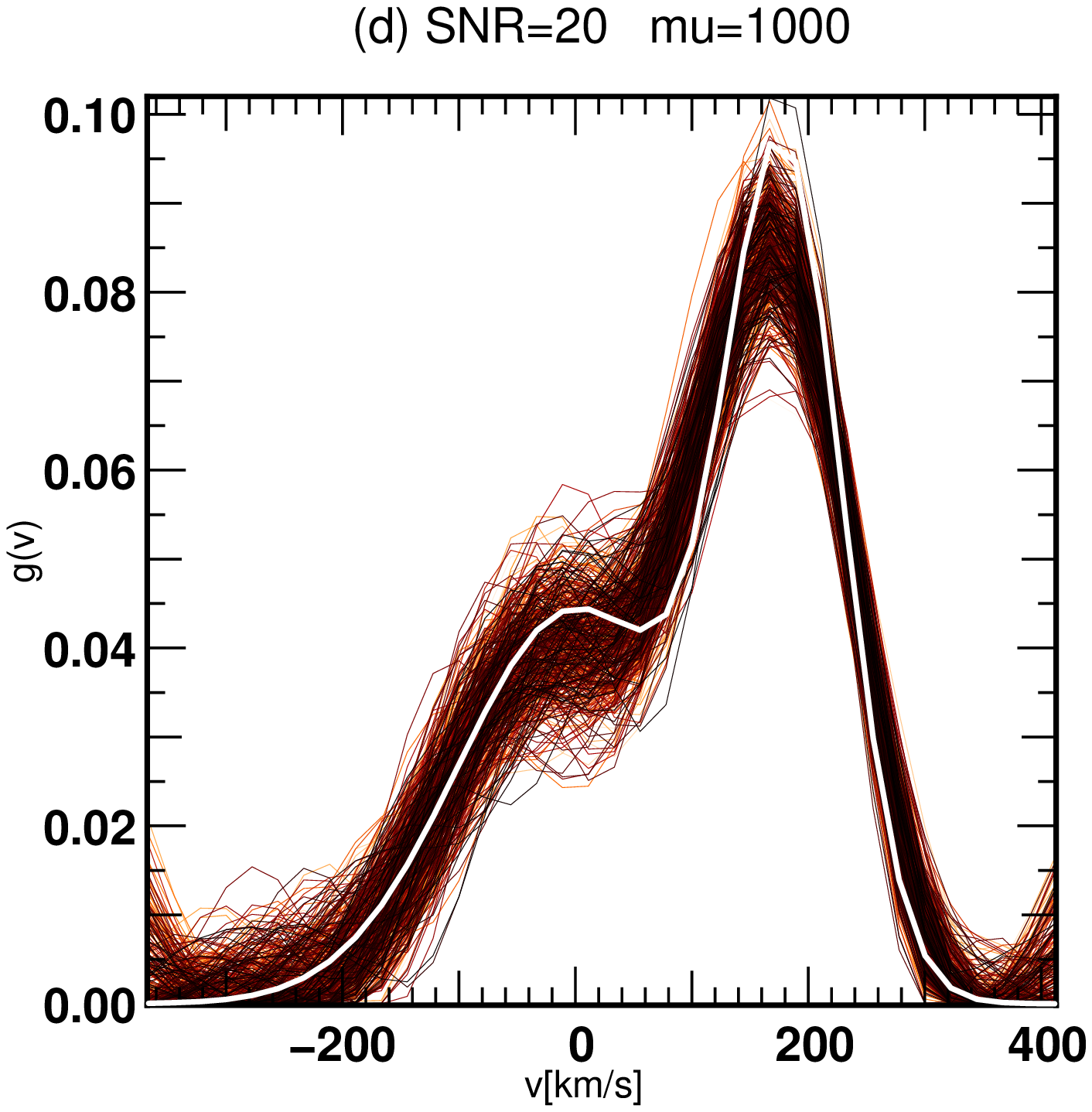} 
\end{tabular}
\end{center}
\caption{
Impact of the smoothing parameter. Reconstruction of a non-Gaussian
LOSVD for simulated data with SNR=20 per pixel.
{\sl Top}: $\mu=10$, {\sl bottom}: $\mu=1000$. {\sl Left}: the thick  line is the 
input model LOSVD. The circles and the bars show respectively the median and
interquartiles of the reconstructed LOSVDs for 500 realizations of the
experiment. {\sl Right}: the whole set of 500 solutions is displayed, with the
model as a thick white line, in order to
give the reader a sense of what individual solutions look like. The figures show the
tradeoff between bias and reliability of the reconstruction: for small $\mu$
the median reconstruction is unbiased but the individual reconstructions are
very noisy. For large $\mu$ the median reconstruction is slightly
biased but all the reconstructions are reasonable. Hence, the significance of an
individual reconstruction is improved by regularization at the cost of introducing a bias.}
\label{f:lossim2}
\end{figure*}

We applied this deconvolution technique to mock data, created from {\PHR }
SSPs of several ages and metallicities, with  $R=10\,000$ at
$4000 - 6800$ \AA. 
In a first set of experiments, the model spectrum at
rest was a solar metallicity $10$ Gyr {\SSP}. It was convolved with various LOSVDs, both Gaussian and non
Gaussian, with velocity dispersions ranging from $30$ to $500$ km/s. It was
then perturbed with Gaussian noise at levels
ranging from SNR$=5$ to $100$ per pixel, and deconvolved using the model
spectrum at rest as template ({\em i.e.} no template mismatch). 
In all cases, the LOSVDs
are adequately recovered. Figure
\ref{f:lossim} shows the reconstruction of a Gaussian LOSVD, for SNR$=10$ per
pixel. However, there are necessarily some biases in the reconstruction of the
sharp features of the LOSVD. This is expected since we introduced
regularization via smoothing.
To illustrate the relationship between regularization and bias, we performed a new set of
similar simulations for a non-Gaussian LOSVD (sum of 2 Gaussians) with SNR=20 per pixel
and varied the smoothing parameter $\mu$. The results are shown in figure
\ref{f:lossim2}. The
panels a and b correspond to $\mu=10$ while the panels c and d correspond to
$\mu=1000$. The model, median and interquartiles of 500 reconstructions are
displayed. We also plotted the whole set of 500 recovered solutions, in order
to show the locus of the solutions. One can see
that the biases of the median recontruction are reduced when lowering
$\mu$. The highest bump is correctly reproduced for $\mu=10$ while it is not
for $\mu=1000$. But on the other hand the solutions are much more widely
spread when $
\mu=10$. This means that most solutions taken from the
set of low $\mu$ simulations can be very far from the model, while all the
large $\mu$ solutions lie reasonably close to the model.

The regularization acts as a Wiener filter in the sense that it damps the high 
frequency components of the solution.  Regularization  improves the 
significance of an individual reconstruction (it will nearly always lie 
reasonably close to the model), at the cost of introducing a bias. 


\subsection{Age and metallicity mismatch}
\label{s:mismatch}
\begin{figure*}
\begin{center}
\begin{tabular}{lll}
\resizebox{5.5cm}{5.5cm}{\includegraphics{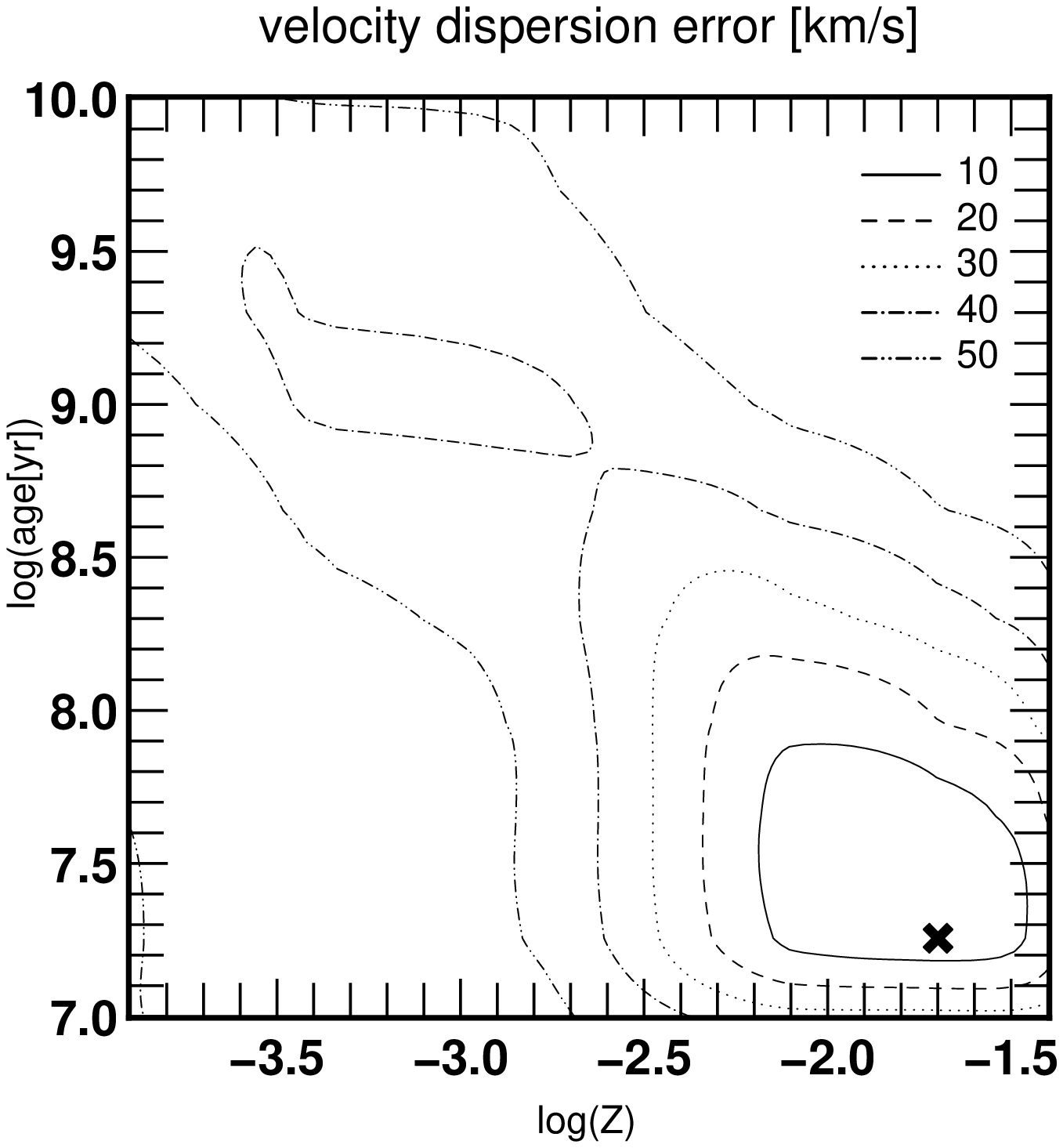}} &
\resizebox{5.5cm}{5.5cm}{\includegraphics{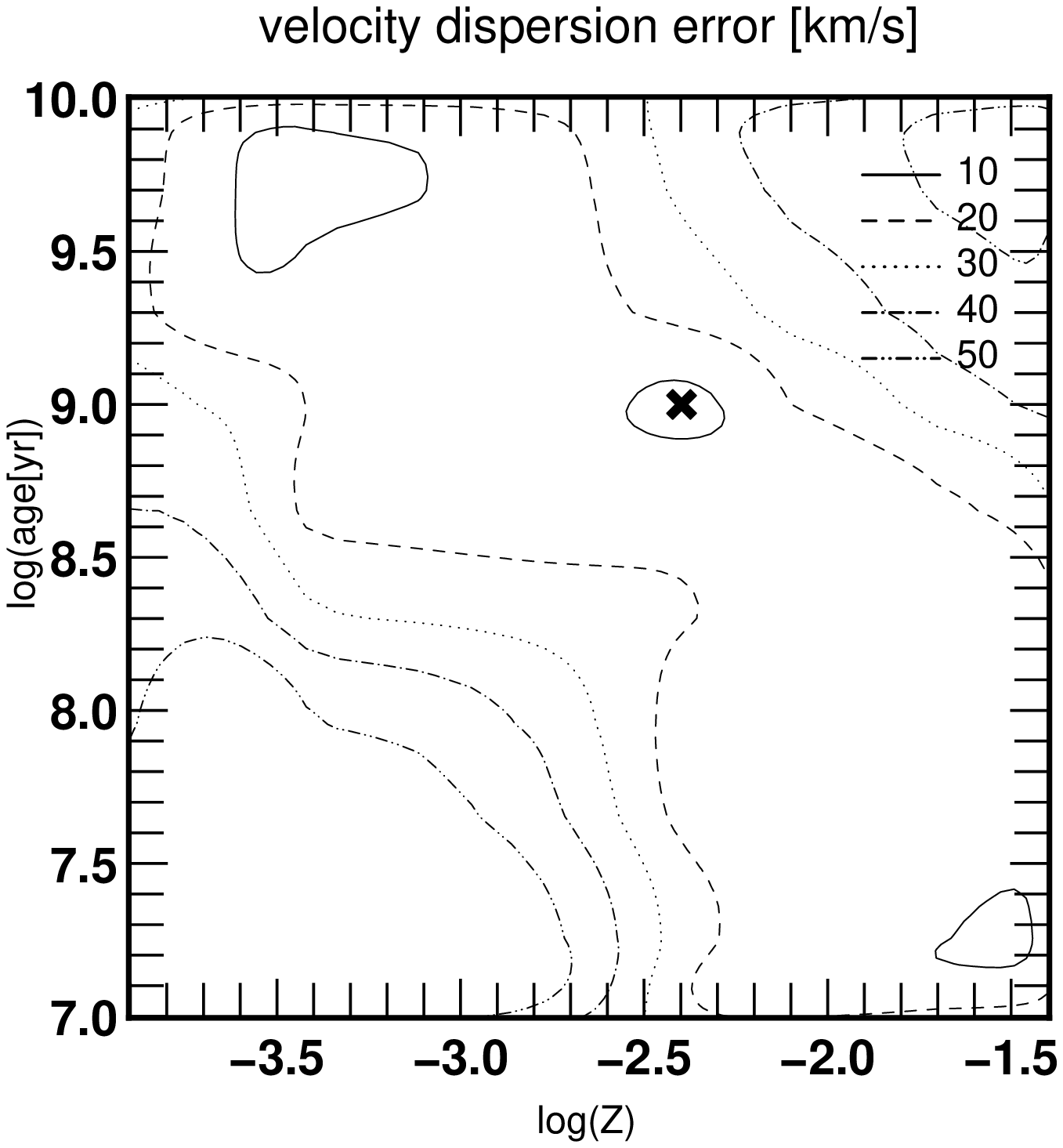}} &
\resizebox{5.5cm}{5.5cm}{\includegraphics{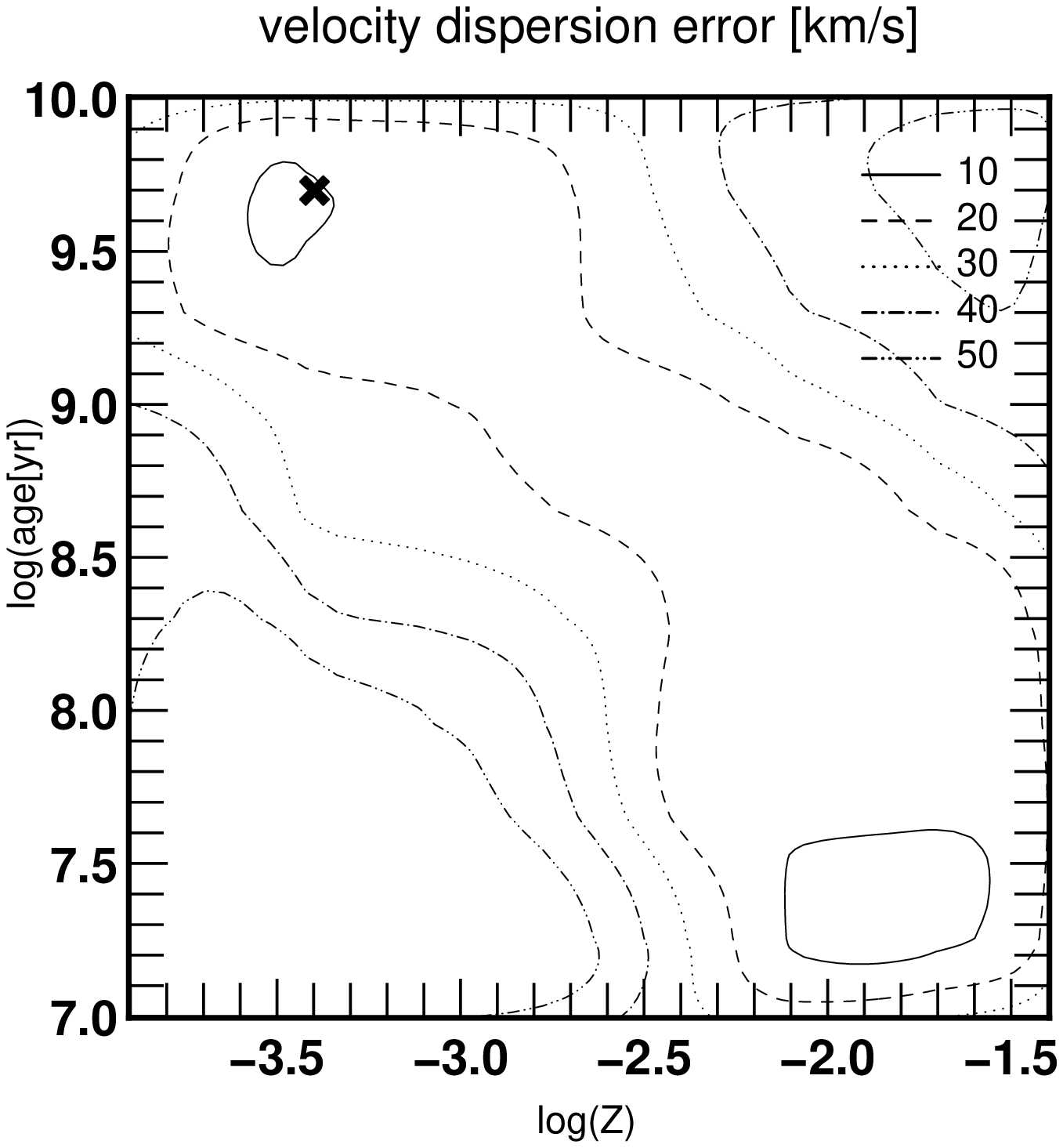}} \\
\end{tabular}
\caption{
Velocity dispersion error as a function of the age and metallicity
of the template \SSP. Contours show regions of increasing velocity dispersion error.
In each experiment, the age and metallicity of the original model template is
shown as a thick cross, and the model LOSVD is a Gaussian with zero mean and
$50$ km/s dispersion. The velocity dispersion error is minimum when the
template's age and metallicity are similar to the model's. The error increases
quickly when the template's parameters differ from the model's, also in the
age-metallicity degeneracy direction (upper left-bottom right diagonal). It increases even faster in the
direction orthogonal to the age-metallicity degeneracy.}
\label{f:TM}
\end{center}
\end{figure*}

We take advantage of the large range of ages and metallicities of {\SSPs}
covered by {\PHR} to shortly illustrate the effects of template mismatch on
LOSVD determinations.
In this section we show the results of a large number of simulations aiming at
characterizing the error made when a wrong template is chosen for the
kinematical inversion of data. For this purpose, mock data were
created by convolving a {\SSP} of age, $a_{0}$, and metallicity, $Z_{0}$, with a
centered Gaussian LOSVD of dispersion $\sigma_v=50$ km/s. It was perturbed
by Gaussian noise corresponding to SNR$=100$ per pixel and then
deconvolved, using as template a {\SSP} of age $a_{1}$, and metallicity
$Z_{1}$. The spectral resolution and wavelength range are the same as in \Sec{sims}.
\Fig{TM} shows the error on the measured velocity dispersion. The latter
is measured as the r.m.s of the reconstructed LOSVD. 
If the parameters of the template are different from those of the model, the
velocity dispersion error increases very quickly. The age metallicity
degeneracy is visible as a valley of smaller error, following the upper-left to bottom-right diagonal of the
figures. 
Of course, the $\chi^2$ distance between the model and the mock data follows a
similar 2D distribution, and will lead to the rejection of highly mismatched
LOSVD estimates. 
{However, in practice, it is usually not straightforward to quantify all the
sources of error. It is thus somewhat arbitrary to set an upper limit of $\chi^2$ for the
admissible solutions, and the error on the
kinematics is thus hard to quantify. This experiment illustrates in this
context the
long known issue that when the
correct model is not available, large errors on the determination of kinematics are expected.}
In order to reduce the error in the estimates of
the kinematical properties of a stellar assembly, it is necessary to allow for
a wide range of modulations of the template. This is
naturally achieved by making the non parametric stellar content account for the changes of
the template, as discussed in the next section.
\section{Recovering stellar content and global kinematics }
\label{s:azek}
The mixed inversion described in this section couples the recovery of both the
stellar content and the kinematics, thereby turning STECMAP into STECKMAP.
Proper application of this method provides an interpretation of the observed object in terms of
stellar content {\em{and}} kinematics.

\subsection{Inverse problem}

For a given model spectrum at rest, $\Frest(\lambda)$, and a LOSVD, $g(v)$, the emitted SED,
$\phi(\lambda)$ is given by \Eq{sk}. We now wish to account also for the additional
variables involved in  $\Frest$, given by \Eq{evergely}, namely the stellar
age distribution, $\Lambda(t)$, the {\AMR} $Z(t)$, and the color excess $E(B-V)=E$.
Injecting \Eq{evergely} into the convolution \Eq{rsk} yields
the emitted SED:
\begin{equation}
\phi(w)=\! \iint \!\! \fext(E,w-u) \Lambda(t) B(w-u,t,Z(t))  g(u)\, \mathd  t\, \mathd u \, ,
\label{e:mi1}
\end{equation}
Solving \Eq{mi1} for $\Lambda$, $Z$, $E$ and $g$ when
$\phi$, $\fext$ and $B$ are given is
the inverse problem we are tackling here. 

\subsection{Discretization and parameters}

Expanding the two time-dependent unknowns $\Lambda(t)$ and $ Z(t)$ as a sum
of $n$ gate functions and injecting into \Eq{evergely} yields the discrete model
spectrum at rest:
\begin{equation} 
{\M{F}}=\diag(\Fext(E)) \cdot \M{B \cdot x} \, ,
\label{e:dmr}
\end{equation}
This discretization is
explained in details in Sec.5 of Paper I.
Similarly, we develop the LOSVD $g(u)$ as a sum of $p$ gate
funtions as in \Sec{npkd}.
Note that the reddened model at rest  plays the role of the stellar template in a
classical kinematic convolution. Injecting \Eq{dmr} into \Eq{conf} thus allows us to express the
model spectrum, $\M{s}$, as 
\begin{equation}
\M{s}= \for \cdot \diag(\fo \cdot \diag(\Fext(E)) \cdot \M{B \cdot x}) \cdot
\fo \cdot \M{g} \, ,
\label{e:azek}
\end{equation}
However, here, the template is  this time modulated by the unknowns describing the
stellar content.

\subsection{Smoothness and metallicity constraints}

The discrete problem of finding the stellar age distribution $\M{x}$, the
{\AMR} $\M{Z}$, the extinction $E$ and the LOSVD $\M{g}$ for an observed
spectrum $\M{y}$ and
given an extinction law $\fext$ and a SSP basis $B$ is of course likely to be very ill-conditioned since it arises as the
combination of several ill-conditioned problems. It therefore requires
regularization. We also want the metallicity of the components to remain in
the model range. We use the standard penalization $P$ and the
binding function $C$ defined in Paper I to build the penalization $P_{\MG{\mu}}$ for this problem:
\begin{eqnarray}
P_{\MG{\mu}}=\mu_{\M{x}} P(\M{x}) +  \mu_{\M{Z}} P(\M{Z})+ \mu_{C}
C(\M{Z}) + \mu_{v} P(\M{g}) \,,
\end{eqnarray}
where ${\MG{\mu}}\equiv (\mu_\M{x},\mu_\M{Z},\mu_{C},\mu_{v})$.
Again, we choose $\M{L=D}_2$ as defined in \citet{pichsieb02}, so that the
penalization $P$ is actually Laplacian. The objective function, $Q_{\MG{\mu}}$, is now defined as:
\begin{equation}
Q_{\MG{\mu}}=\chi^2 (\M{s}(\M{x,Z,}E,\M{g})) + P_{\MG{\mu}} (\M{x,Z},E,\M{g}) \, .
\end{equation}
and its partial derivatives are given in \Sec{gradazek}.
Note that there is in principle an additionnal formal degeneracy for this inverse problem. If the set $(\M{x,Z},E,\M{g})$ is a solution to
(\ref{e:mi1}), then $(\alpha \M{x,Z},E,\M{g}/\alpha)$ is also a solution for
any scalar $\alpha$,
because the age distribution $\M{x}$ and the LOSVD $\M{g}$ are not explicitly
normalized in this formulation. However, the adopted regularization 
lifts this degeneracy. The penalization function $P$ is quadratic  ($
P(\alpha\M{x})=\alpha^{2}P(\M{x})$). Thus, if $\M{x}$ or $\M{g}$ is too
large in norm,
the solution is unattractive. Practically, the algorithm reaches a solution
where $\M{x}$ and $\M{g}$ are similar in norm. In any case, this degeneracy
would easily be
remedied by adding a normalizing term to the penalization $P_{\MG{\mu}}$ of the form
$\Vert \M{x} \Vert -1$, which would force the discretized stellar age
distribution $\M{x}$ to have unitary norm. Following the same principle, one could equivalently choose to
normalize the LOSVD rather than the {\SAD }.
\subsection{Simulations}

\begin{figure*}
\begin{tabular}{lll}
\resizebox{5cm}{5cm}{\includegraphics{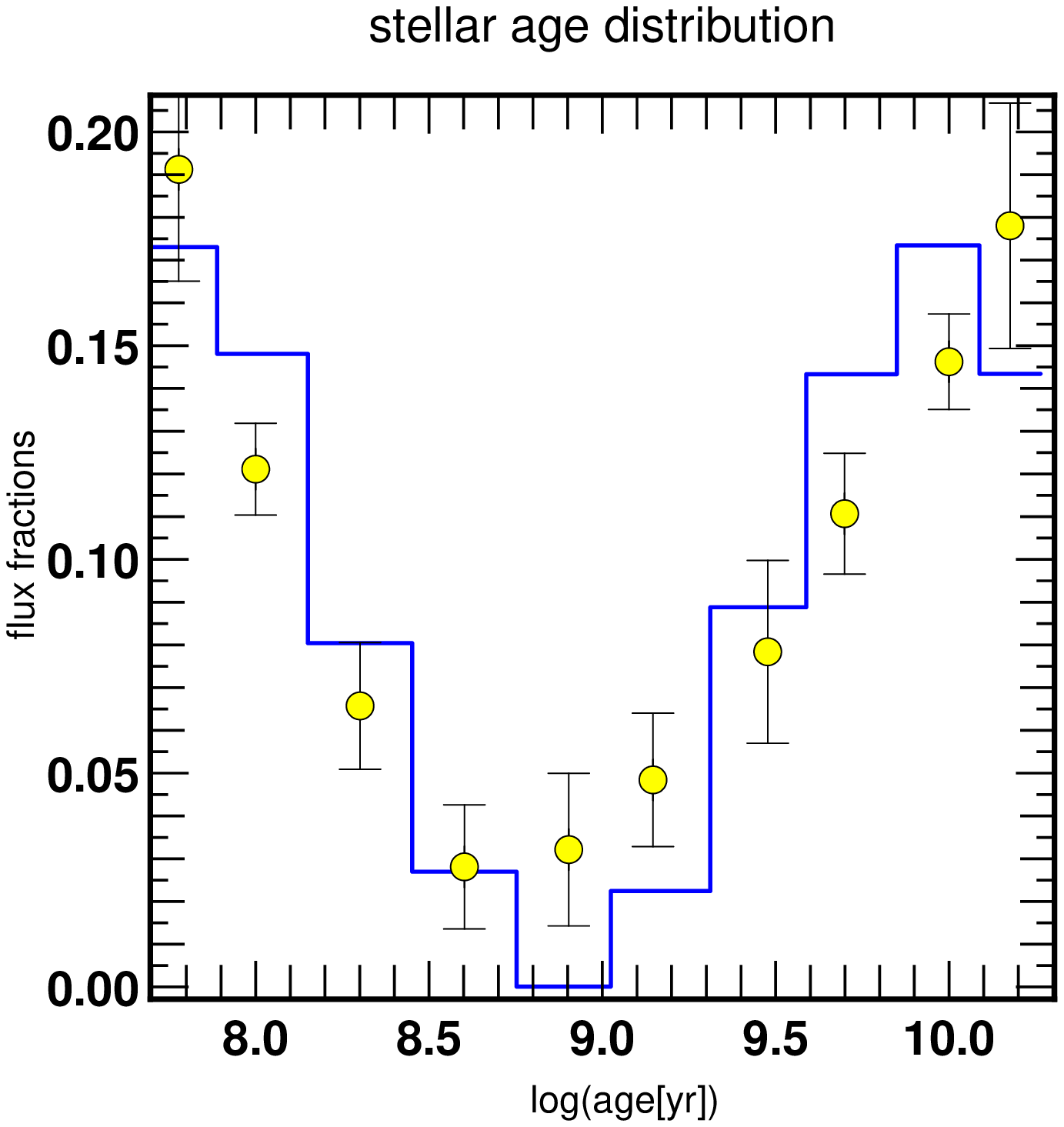}} &
\resizebox{5cm}{5cm}{\includegraphics{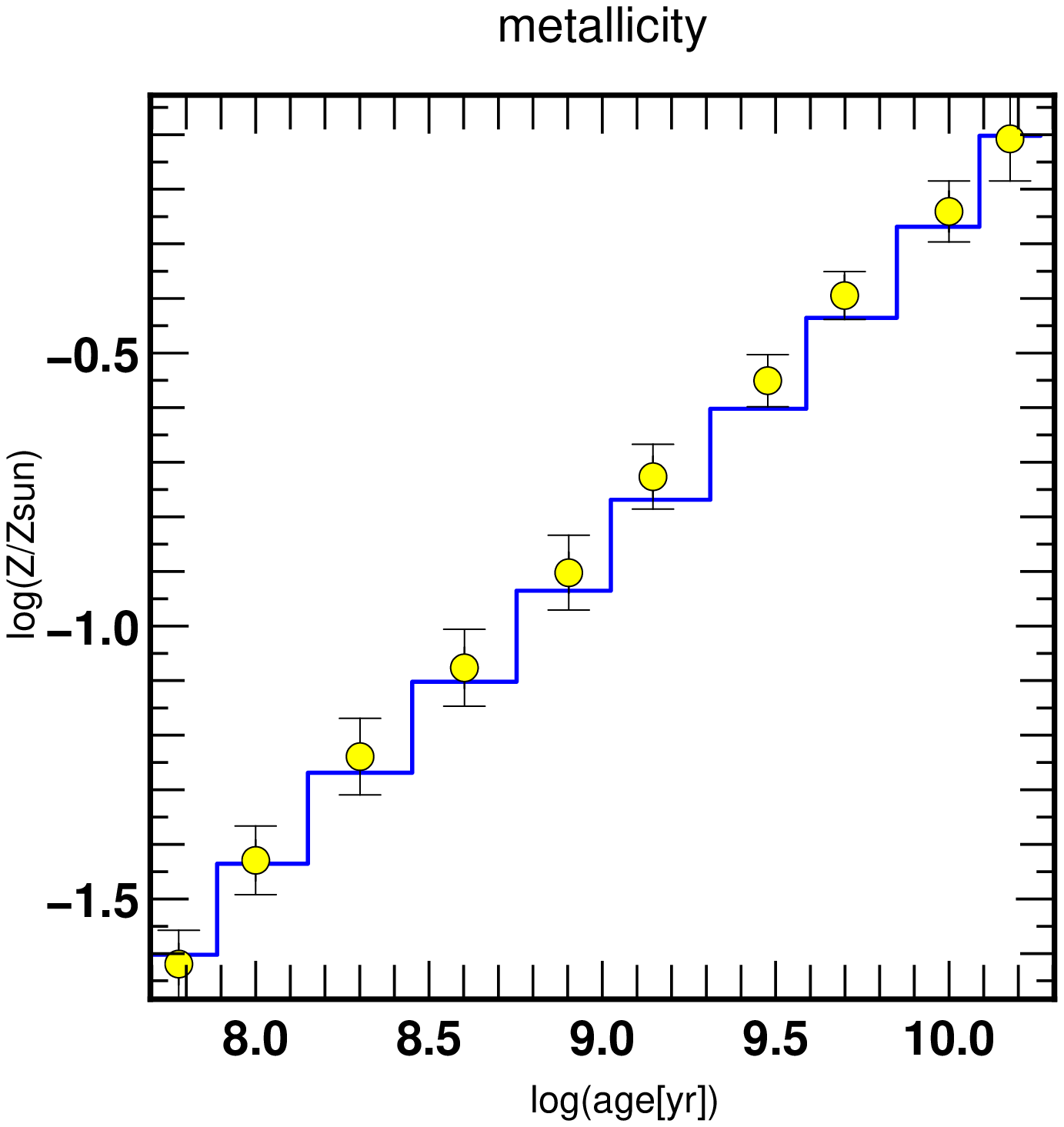}} &
\resizebox{5cm}{5cm}{\includegraphics{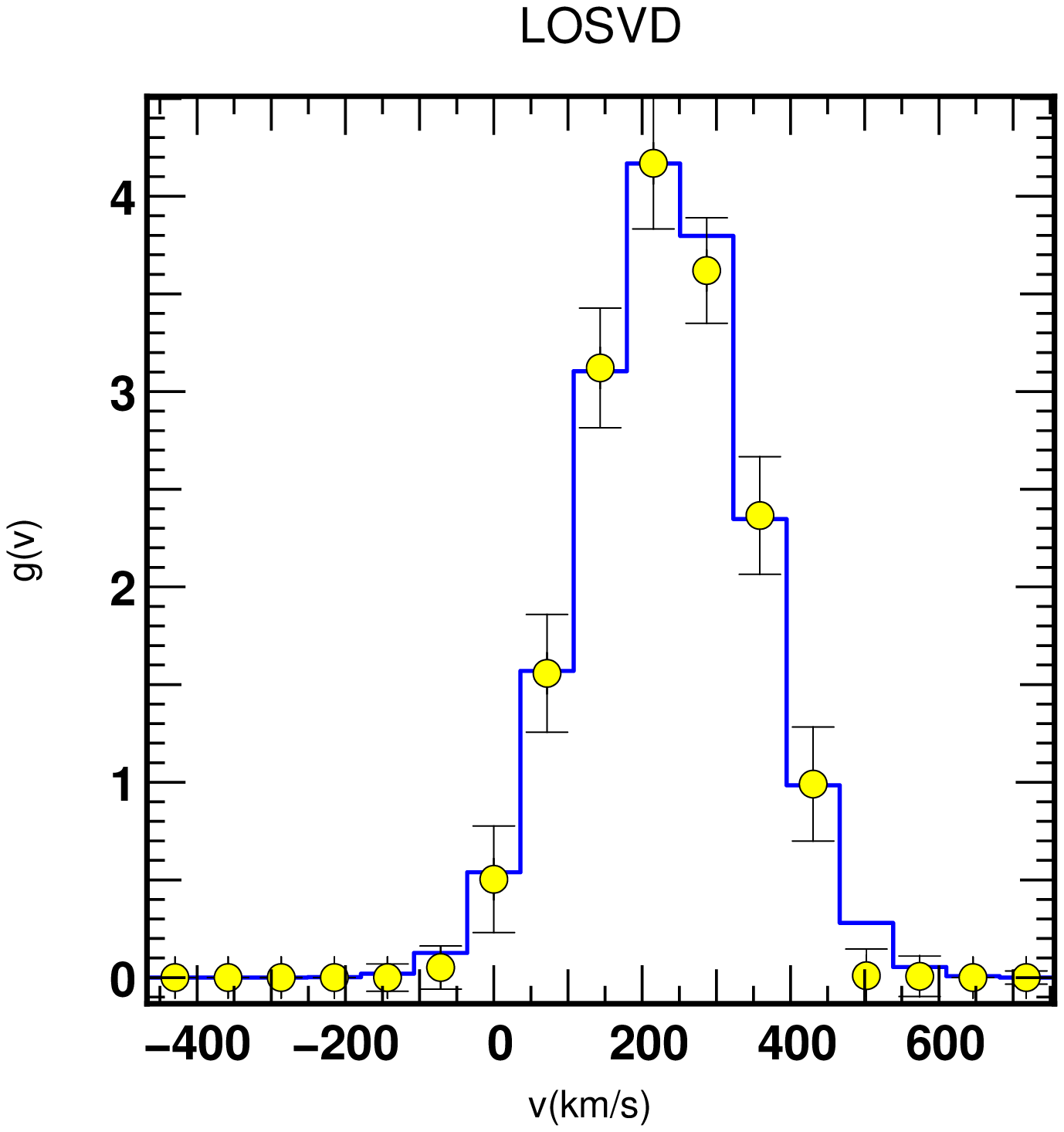}} \\
\resizebox{5cm}{5cm}{\includegraphics{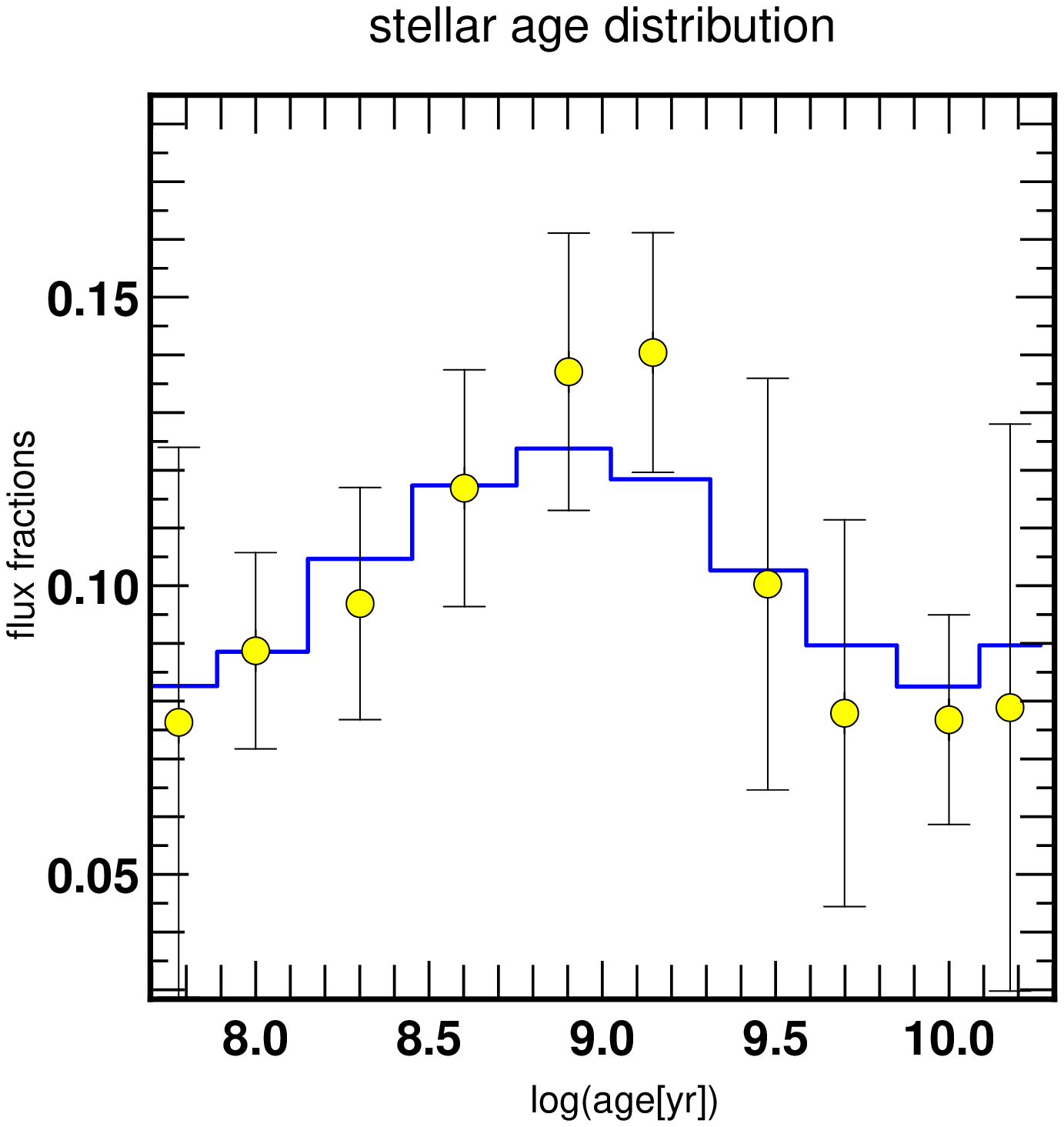}} &
\resizebox{5cm}{5cm}{\includegraphics{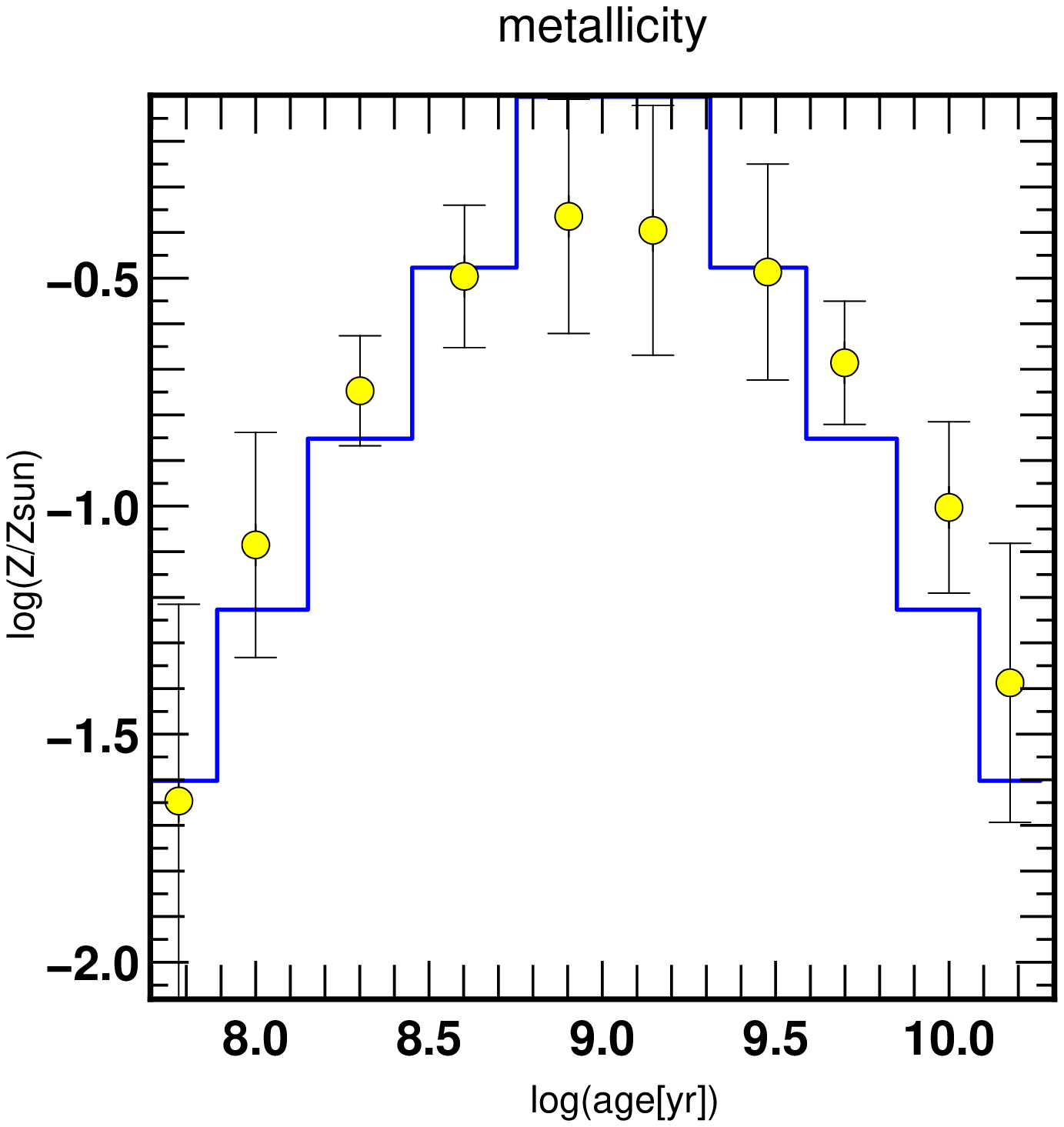}} &
\resizebox{5cm}{5cm}{\includegraphics{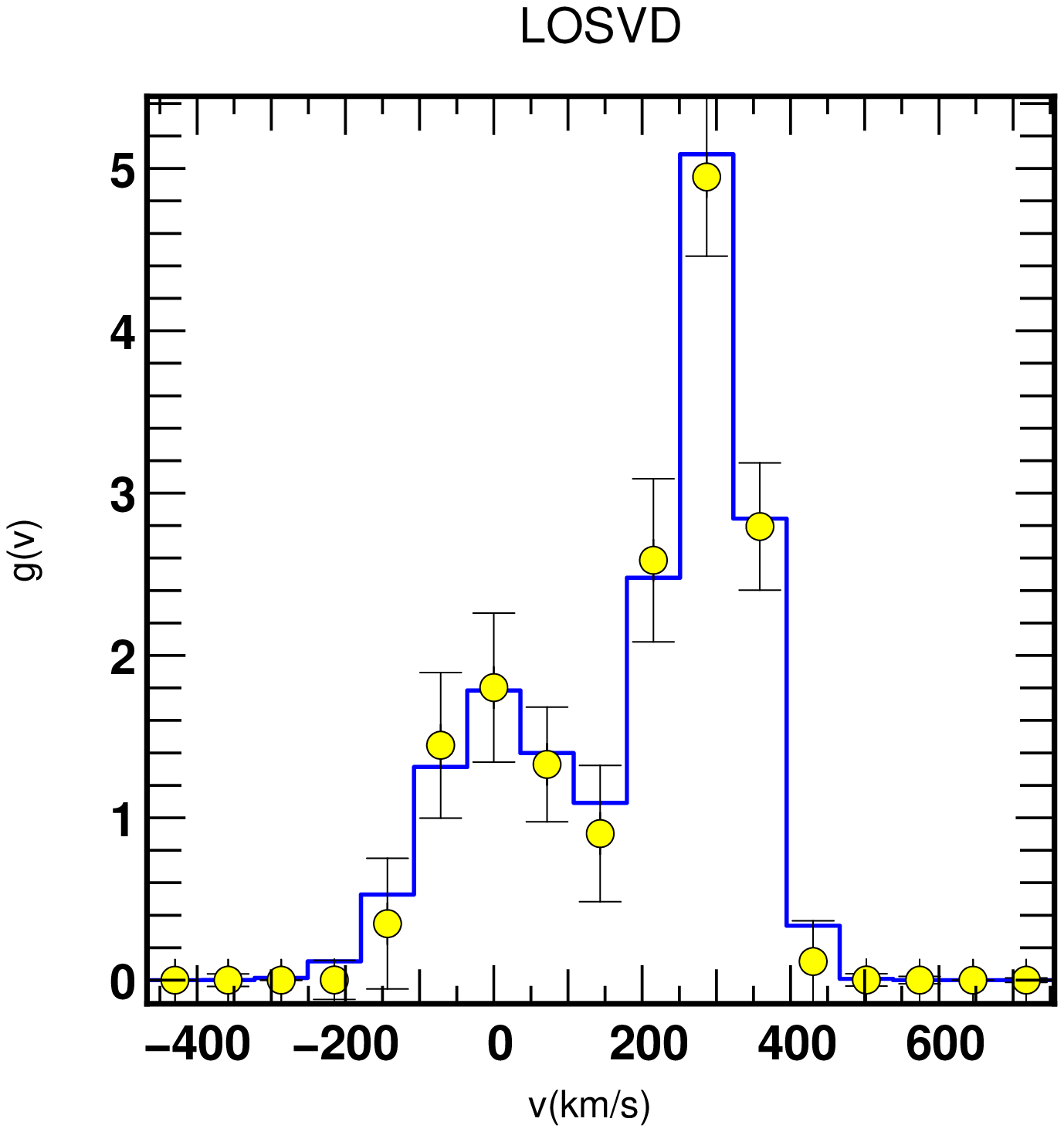}} \\
\end{tabular}
\caption{Reconstruction of the stellar age distribution, {\AMR} and
LOSVD for simulated SDSS-like data with SNR=30 per pixel. The thick histogram
is the input model. The circles and the bars show respectively the median and the
 interquartiles of the recovered solutions for 50 realizations.}
\label{f:azeksim}
\end{figure*}

Let us now test the behaviour of STECKMAP by applying it to mock data.
The latter were produced using an arbitrary stellar age distribution
$\M{x}$, an {\AMR} $\M{Z}$, a LOSVD $\M{g}$ and an extinction parameter $E$. Several simulations were performed with various
input models: bumpy age distributions, increasing or decreasing {\AMR} and
extinctions, Gaussian and non Gaussian wide or narrow LOSVDs, in various
pseudo-observational contexts.
Figure \ref{f:azeksim} shows the results of two of these experiments. 
In the top line, the model is a young metal-poor population
superimposed to an older metal-rich population. In the bottom panels, the
model has a rather constant {\SAD}, a non-monotonic {\AMR}
and a strongly non Gaussian LOSVD. In both cases the 3 unknowns are correctly
recovered. 
In these examples, the data quality mimics that of the best Sloan Digital Sky
Survey galaxies: the resolution is $R \approx 2000$ and SNR$=30$ per
$\approx$1 {\AA } pixel. The wavelength domain of {\PHR} is however narrower than the SDSS's. 
These simulations simply aim at demonstrating the generally good behaviour of the
method, and show that accounting for the kinematics does not fundamentally weaken the
constraints on the stellar content. For a more thorough study of the informational content of the {\PHR}
wavelength range, the reader can refer to the systematic double burst
simulations with variable spectral resolution and SNR per {\AA} performed in
Paper I.


\section{Recovery of age-dependent kinematics}
\label{s:agekin}
%
\label{s:adk}
In this section we present an implementation of the recovery of age-dependent
kinematics, i.e the situation when each sub-population has its {\sl own} LOSVD. In this
experiment, we restrict ourselves to the case where the stellar
populations have a known metallicity and are seen
without extinction. This choice is mainly motivated by the numerical cost of
such a large inversion procedure.
The modeling is given by \Eq{ckrs}. Finding
the age-velocity distribution $\V(u,t)$ when the mono-metallic basis $B$ and the observed
spectrum $\phi$ are given is the inverse problem. It arises as the
combination of a linear age inversion and a kinematical deconvolution.
%
%
\subsection{A sum of convolutions}

The age-velocity  distribution, $\V(u,t)$, is  expanded as a
linear   combination  of   normalized   2D  gate   functions
$\theta_{ij}(u,t)$:
\begin{displaymath}
  \theta_{ij}(u,t) \equiv \frac{1}{\delta{}u\,\delta{}t} \,
	\theta\left(\frac{u - u_{i}}{\delta{}u}\right)\,
	\theta\left(\frac{t - t_{j}}{\delta{}t}\right)\,.
\end{displaymath}
In  other words,  $\V(u,t)$  is represented  by  a 2D  array
$\M{v}$ of size $(p,n)$, {\em  i.e.} $p$ is the size of each
LOSVD and $n$ is the number  of age bins. The linear step in
$u$ is $\delta u$ and the step in $t$ is $\delta t$.

%
By injecting the expansion into \Eq{ckrs} we obtain
\begin{eqnarray}
\phi(w) &=& \iint\sum_{i=1}^{p} \sum_{j=1}^{n}   v_{ij} \theta_{ij}(u,t) B(w-u,t)
\mathd t \mathd u \, ,  \nonumber \\
&\simeq&  \sum_{i=1}^{p}
\sum_{j=1}^{n} v_{ij}  B_{j}(w-u_{i}) \, .
\end{eqnarray}
As in the previous sections, $B_{j}(u)$ is a time-averaged {\SSP}
of age $t_{j}\pm\frac{1}{2}\,\delta{}t$. 
We then discretize along wavelengths by averaging over small $\delta w$:
\begin{eqnarray}
  \phi_{k}&=&
    \frac{1}{\delta w} \sum_{i=1}^{p}
    \sum_{j=1}^{n} v_{ij} \, \int B_{j}(w - u_{i})\,
    \theta\left(\frac{w - w_{k}}{\delta\,w}\right)
    \mathd w  \, , \nonumber  \\
  &\simeq&
    \sum_{i=1}^{p}\sum_{j=1}^{n}
    v_{ij}\,B_{j}(w_{k}-u_{i}) \, , 
\end{eqnarray}
where  $(w_{j})_{j \in
\{0,...,m \} }$ is a set of constant step logarithmic wavelengths.
The above expression also reads in matrix form as a sum of kernel
convolutions. Finally, the model SED of the emitted light reads:
\begin{equation}
\M{s}=\sum_{j=1}^{n} \M{K}_{j}\cdot \M{v}_{j} \, ,
\label{e:cksk}
\end{equation}
where $\M{s}=\left( \phi_{1} , \phi_{2} , \dots , \phi_{m} \right)$,
$\M{v}_{j}=\left(  v_{1j},  v_{2j} , \dots , v_{pj} \right)$ and
\begin{equation}
\M{K}_{j}=\left[
\begin{array}{rrrr}
 K_{11j} & K_{12j} & \dots &  K_{1pj} \\ 
K_{21j} & K_{22j} & \dots & K_{2pj} \\ 
 \vdots & \vdots &  \ddots & \vdots \\
K_{m1j} & K_{m2j} & \dots & K_{mpj}
\end{array}
\right] \, ,
\end{equation}
with
\begin{equation}
K_{ikj} \equiv   B_{j}(w_{k}-u_{i}) \, .
\end{equation}
With this notation, $\M{K}_{j}$ and $\M{v}_{j}$ are respectively the convolution kernel and the LOSVD of the sub-population of age $t_{j}$, and the
model spectrum $\M{y}$ is the sum of the convolution of the kernel of each
sub-population by its own LOSVD. 
%
\subsection{2D age-velocity smoothness constraints}
In the previous sections, the unknowns were mono-dimensional functions of time or velocity. Here, the unknown is a 2D distribution,
and we thus have to implement a 2D smoothing constraint. We
wish to allow the smoothness in age to be distinct from the smoothness in
velocity. We thus construct two penalizing functions, $P_{a}$ and $P_{v}$, relying on the
standard function $P$. $P_{a}$ computes the sum of the Laplacians of the columns of $\M{v}$ while $P_{v}$ computes the sum of the Laplacians of the lines of $\M{v}$.
The smoothness in the direction of the velocities (respectively ages) is set
by $\mu_{v}$ (respectively $\mu_{a}$). We define the
vectors $\M{v}_{j}=(v_{1j}, v_{2j}, \cdots, v_{pj})$ as the columns of
$\M{v}$, {\em i.e.} the LOSVD s of the subpopulations. We similarly define the $\M{v}^{i}=(v_{i1}, v_{i2}, \cdots,
v_{in})$ as the lines of $\M{v}$. With this notation, the penalization
$P_{\MG{\mu}}$ reads:
\begin{eqnarray}
P_{\MG{\mu}} ({\M{v}}) &\equiv& \mu_{a} P_{a}({\M{v}}) + \mu_{v} P_{v}(\M{v}) \, ,
\nonumber \\
&\equiv&
\mu_{a} \sum_{i=1}^{p} P({\M{v}}^i) + \mu_{v} \sum_{j=1}^n P({\M{v}}_{j})  \, .
\end{eqnarray}
The objective function, $Q_{\MG{\mu}}$,
 is now fully specified as $Q_{\MG{\mu}}= \chi^2 +
P_{\MG{\mu}}$. Its gradients are given in appendix \ref{s:aadk}. 
We choose here the
 smoothing parameters, $\MG{\mu}\equiv (\mu_{a},\mu_{v})$, on the basis of simulations.
\subsection{Simulations of a bulge-disk system}
\begin{figure*}
\begin{center}
\resizebox{7cm}{7cm}{\includegraphics{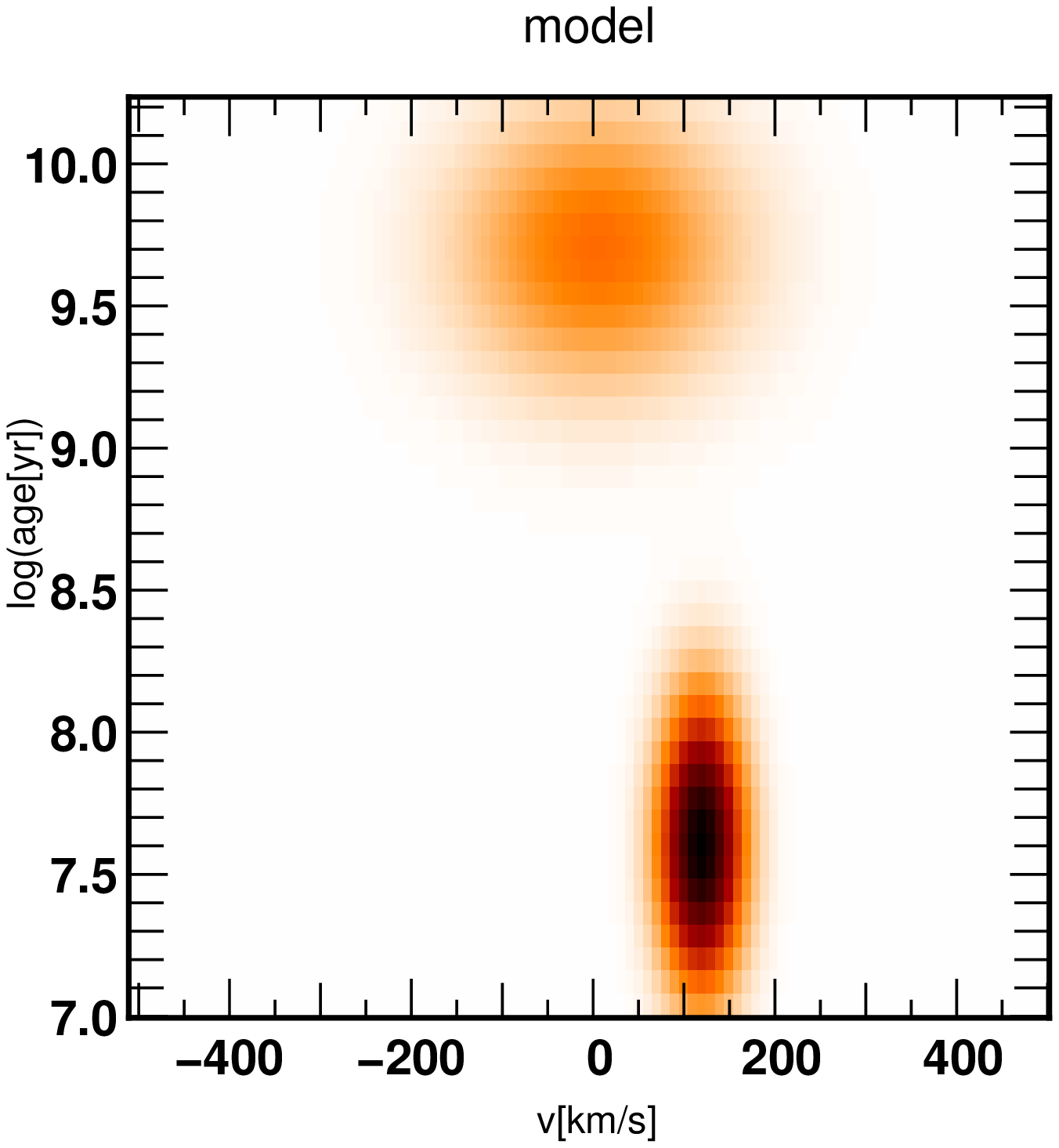}} 
\resizebox{7cm}{7cm}{\includegraphics{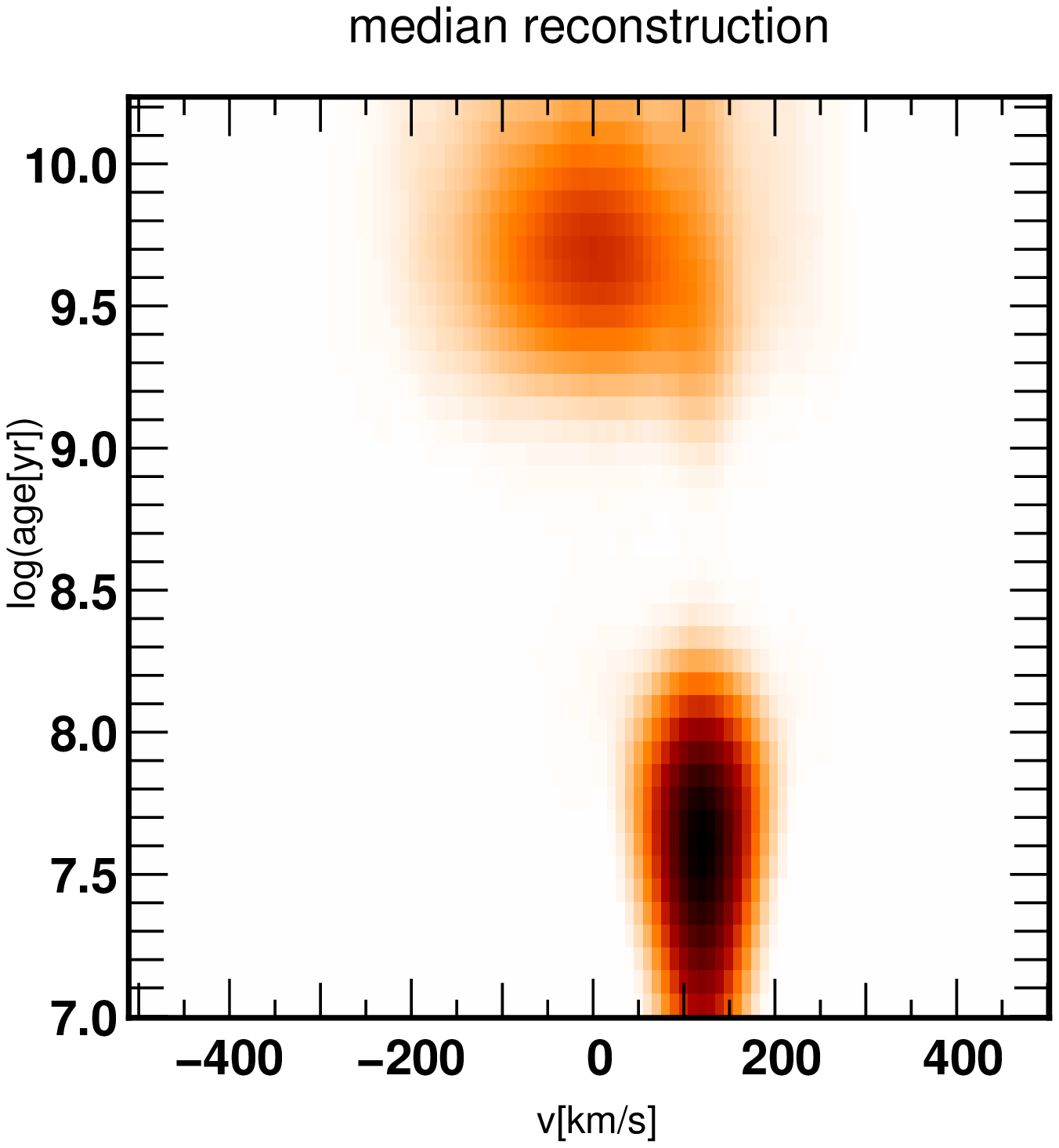}} 
\end{center}
\caption{Model (left) and median reconstruction (right, $\approx$ 30
realizations) of a stellar age-velocity distribution from SNR$=$100
per $0.2$ {\AA } pixel pseudo-data at 4000$-$6800 {\AA}. The model stellar age-velocity distribution mimics that
of a simplified spiral galaxy seen with intermediate inclination. The old broad
component can account for the bulge population while the young narrow rotating
component represents a thin disk population. The projected kinematical parameters of the
model are given in \Tab{bdsim} (case 1). The different kinematical
components are well separated and clearly identifiable.}
\label{f:simak}
\end{figure*}

\begin{figure*}
\begin{center}
\resizebox{7cm}{7cm}{\includegraphics{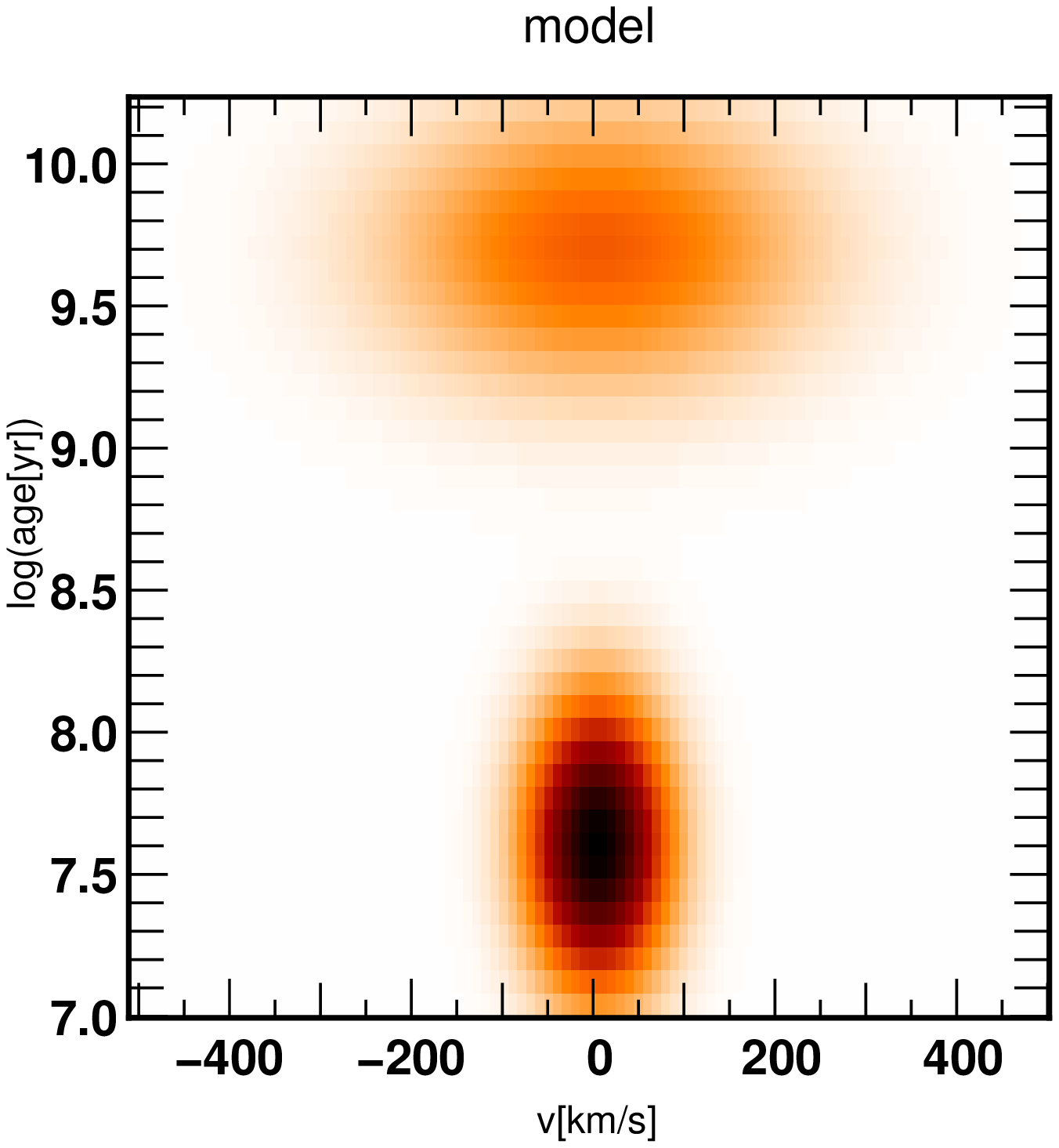}} 
\resizebox{7cm}{7cm}{\includegraphics{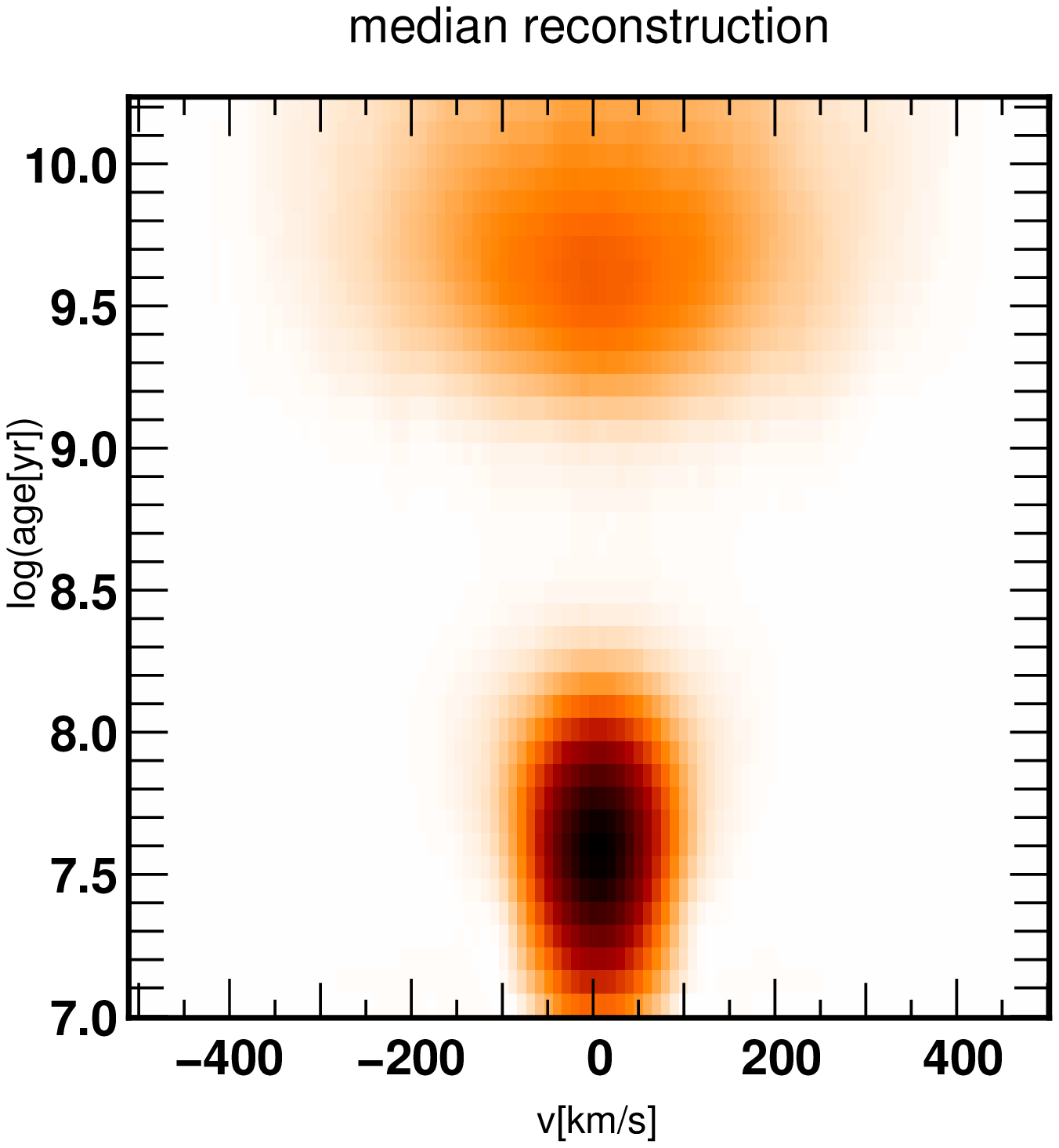}} 
\end{center}
\caption{Same as \Fig{simak} but for an unresolved simplified spiral galaxy with
projected and spatially integrated kinematical parameters given in \Tab{bdsim}
(case 2). The velocity dispersion of the integrated young disc component
depends on the inclination angle. The bulge and the disc are are well
separated and clearly identifiable. Their respective velocity dispersions and
ages are
reliably recovered.}
\label{f:simak2}
\end{figure*}

We studied the feasibility of
separating two age-dynamically distinct populations, {\em i.e.} two components
which do not overlap in an age-velocity distribution diagram, in a regime of  very high
quality model and data.
We performed simulations  in the idealized case
of a very simplified spiral galaxy consisting of a bulge-disk system of solar metallicity seen without extinction at
some intermediate inclination, in two observational contexts. The
corresponding ages and projected kinematical parameters are given in \Tab{bdsim}.  The resolution of the pseudo-data is $R=10\,000$ at $4000-6800$ \AA, and the SNR
is $100$ per $0.2$ {\AA} pixel.
\begin{itemize}
\item{ {\em Case 1:} The galaxy is resolved, and the fiber aperture is small
compared to the angular size of the galaxy. The line of sight is offset by a couple kpc from the center
along the major axis.
The projected model age-velocity distribution involves 2 superimposed
components: an old, non-rotating kinematically hot
population representing the bulge, and a young, rotating,
kinematically cold component. The model and the median of 30 reconstructions are shown in \Fig{simak}.
The separation of the components is clear and their
parameters can be recovered with good accuracy, considering the difficulty of
the task.
}
\item{ {\em Case 2:} The galaxy is unresolved. 
The difference with the former situation is that because of the spatial
integration, both age-velocity distributions are centered. For a given
dynamical model, the projected dispersion of the disk component depends on its
inclination. \Fig{simak2} shows that the separation is successful and
that the ages and integrated kinematical properties of both components can be
measured.}
\end{itemize}
\begin{table}
\begin{center}
\begin{tabular}{c|c|c|c}
& $V_{c} (\rm{km/s})$ & $\sigma_{\rm{v}} (\rm{km/s})$ & \rm{age (Gyr)} \\
\hline
\rm{Case 1} \\
\hline
\rm{Bulge} & 0 & 100 & 8  \\
\rm{Disk} & 120 & 30 & 0.5
 \\
\hline
\rm{Case 2} \\
\hline
\rm{Bulge} & 0 & 150 & 8  \\
\rm{Disk} & 0 & 50 & 0.5
\end{tabular}
\end{center}
\caption{Projected kinematical parameters and age of the model bulge-disk system used to
produce the simulations of \Fig{simak} and \Fig{simak2}. $V_c$ (respectively
$\sigma_{\rm{v}}$) is the rotation velocity (respectively the velocity
dispersion) 
projected on the line of sight.}
\label{t:bdsim}
\end{table}
%
%
%
%
%
%
%
\section{Conclusions}
\label{s:conclusion}
The non-parametric kinematical deconvolution of a galaxy spectrum
is efficiently performed using a MAP formalism (\Sec{kin}). Regularization through
smoothness requirements and positivity improve significantly the behaviour of
the inversion with respect to noise in the data. This improvement occurs
at the cost of introducing some bias in the reconstructed LOSVD, but this bias
remains reasonable. Strong non-Gaussianities of LOSVDs are reliably detected from
mock data generated using {\PHR} SSPs for SNR down to $20$ per $0.2$ {\AA } pixel.

When the template does not exactly match the model spectrum at rest,
i.e. there is some template mismatch, the error on the velocity dispersion
increases very quickly (\Sec{mismatch}). {For example, in our
experiments, where $\sigma_{\rm{V}} = 50$ km/s with $R=10\,000$ data, the error on
the measured velocity dispersion amounts up to 10-20 per cent if the template
differs from the model by more than 0.3 dex in age and metallicity, perpendicular to
the age-metallicity degeneracy.} 

The formal similarity between the non-parametric kinematical
deconvolution and the recovery of the stellar content allows us to merge both
processes in a ``mixed'' inversion where the observed spectrum is fitted by determining the
stellar content and the kinematics simultaneously (\Sec{azek}). This circumvents the need
for iterations  where kinematical and stellar content analyses are
carried out one after the other, until convergence is reached; this provides an
efficient method to analyze large sets of data.

Satisfactory reconstructions of the {\SAD}, the {\AMR}, the
extinction and the global LOSVD were obtained from mock data down to $R=2000$,
SNR$=30$ per 1 {\AA} pixel in the $4\,000 - 6\,800$ {\AA} range (simulating SDSS data
in the {\PHR} range), 
indicating the good behaviour of the method. 
Since in our simulations, the introduction of the kinematics into
STECMAP did not affect the recovery of the stellar content, we
consider that the
error estimates and separability analysis given in Paper I remain
valid.

%
%
In a more exploratory part of this work, we showed the feasibility of  recovering age-dependent kinematics
in a simplified mono-metallic unreddened context (\Sec{agekin}). We were able to separate the bulge and disc component of a simplified model
spiral galaxy in integrated light provided very high quality data
(SNR=100 per $0.2$ {\AA} pixel in the optical domain) and
models are available, {\em i.e.} we constrain both components in velocity
dispersion and age. This separation was also carried out successfully in the
setup corresponding to an unresolved galaxy. 

Further investigations are needed to extend this
technique to a regime where the metallicity and extinction are unknown.
We expect that letting the metallicity be a free parameter would certainly lead to a more
degenerate problem, as shows the degradation of the resolution in age found in
Paper I compared to fixed metallicity problems. On the contrary, we
do not expect the addition of the extinction as a free parameter or a more complex
form of extiction law or flux calibration correction, possibly non parametric,
to deteriorate the conditioning of the problem.
The results are encouraging, and the
feasibility of such age-dependent kinematics reconstructions deserves to be tackled
in realistic specific pseudo-observational regimes in the future.


As mentioned in Paper I, the SSP  models were considered to be
perfect and noiseless. It still has to be investigated how instrumental error
sources such as flux and wavelength calibration error, additive noise, 
contamination by adjacent objects, and, equally important, model errors, can
affect the robustness of such sophisticated interpretations. 

\section*{Perspectives}
\label{s:persp}
STECKMAP will be very useful to interpret data of large spectroscopic surveys, complete or
in progress, such as 2DFGRS\footnote{http://www.mso.anu.edu.au/2dFGRS/}, SDSS\footnote{http://www.sdss.org/},
DEEP2\footnote{http://www.deep.berkeley.edu/}, or
VVDS\footnote{http://www.oamp.fr/virmos/vvds.htm}, especially
where both constraints on the stellar content and the dynamics are required. 
STECKMAP's analysis of the spectroscopic survey data or of a SNR
selected subsample, combined with 
survey photometry could
provide  estimates of the stellar and dynamical masses (which must be
corrected for fiber aperture though), thereby allowing astronomers
the prospect of  investigating the dark
matter content in galaxies on a statistically significant sample, in the
spirit of \citet{padmanabhan04}.

The application of age-dependent kinematics to integral field spectroscopy
data from, for example SAURON \citep{bacon01b,dezeeuw02}, OASIS
\citep{mcdermid04}, MUSE \citep{muse03} or MPFS \citep{chilingarian04} could significantly boost the
amount of information extracted from this data.

The inner parts of elliptical or dwarf
elliptical galaxies have shown via adaptive optics new kinematically
decoupled structures (cores or central disks), which were precedently
unresolved \citep{mcdermid04-1,bacon01a}. 
Similarly, if decoupled structures are unresolved and remain so, even with
adaptive optics, it may still be possible to separate components in age-velocity
space. Hence, the technique presented in \Sec{agekin} extends the range of investigation  for
the inner components of galaxies even further in redshift and distance with the
current generation of instruments.
The faint,  generalized  counterparts of kinematically decoupled cores, {\em
i.e.} stellar streams generated by minor merging and accretion of satellites,
may also be detectable by an age-dependent kinematics reconstruction in systems which can not be resolved into stars,
provided that they are sufficiently distinct from the bulk stars of the
galaxy in
the age-velocity space. This will enlarge the sample of galaxies
for which such detailed information is available, and may make it
statistically significant.

{\bf Acknowledgments}

 {\sl  {We   are  grateful   to  A.~Siebert for  useful comments
 and  helpful  suggestions}.  We  would  like  to  thank D.~Munro  for  freely
 distributing   his  Yorick  programming   language  (available   at  {\em\tt
 http://www.maumae.net/yorick/doc/index.html}),   together  with   its   {\em\tt  MPI}
 interface,  which  we used  to  implement  our  algorithm in  parallel.  
PO thanks the MPA for their hospitality and funding from a Marie Curie
studentship. }


\bibliographystyle{mn2e}
\bibliography{STECKMAPbib}

\vfill


\appendix

\section{Gradient computations}

\subsection{Kinematic deconvolution}

\label{s:gradkd}

In this section we derive the gradient of $Q_{\mu}$ with respect to the LOSVD
$\M{g}$. 
First, we rewrite the $\chi^2$ term as:
\begin{equation}
\chi^2=\T{\M{r}} \cdot \M{W} \cdot \M{r}\, ,
\end{equation}
where the residuals vector $\M{r}$ is defined by 
\begin{equation}
\M{r}=\M{y}-{\fo}^{-1} \cdot \diag (\fo \cdot \M{F}) \cdot \fo \cdot \M{g}\, ,
\end{equation}
The derivative of the $\chi^2$ then reads:
\begin{equation}
\pdrv{\chi^2}{\M{g}}= -2 \for \cdot \diag(\fo \cdot \M{F})^\ast \cdot \fo
\cdot \M{W
\cdot r} \, ,
\label{e:drsk}
\end{equation}
where the asterisk $\ast$ denotes the complex conjugate.
Since the stellar template and the LOSVD can play symmetrical roles in
\Eq{conf} we can also write the derivative of ${\chi^2}$ relatively to the stellar
template:
\begin{equation}
\pdrv{\chi^2}{\M{F}}= -2 \for \cdot \diag(\fo \cdot \M{g})^\ast \cdot \fo
\cdot \M{W
\cdot r} \, ,
\label{e:drs}
\end{equation}
This expression will be useful
for later derivations of gradients for more complex problems in the following appendices.

\subsection{Gradients of the mixed inversion}

\label{s:gradazek}

Here we show how to obtain the partial derivatives of $Q_\mu=\chi^2 + P_\mu$
as defined in \Sec{azek}. Given that writing the derivatives of the penalizing
functions $P_\mu$ is
straightforward, we will in this appendix focus on the
gradients of the $\chi^2$.
In the mixed inversion, the reddened
model spectrum at rest plays the role of the stellar template $\M{F}$ in the classical
kinematic deconvolution of equation (\ref{e:mk}). $\partial \chi^2 / \partial
\M{g}$ can thus be obtained by replacing $\M{F} \leftarrow   \diag(\Fext(E)) \cdot \M{B \cdot x}$ in equation (\ref{e:drsk}).
\begin{equation}
\pdrv{\chi^2}{\M{g}}= -2 \for \cdot \diag(\fo \cdot \diag(\Fext(E)) \cdot \M{B
\cdot x} )^\ast \cdot \fo
\cdot \M{W \cdot r} \, ,
\end{equation}
where $\M{r}={\M{y-s}}$ is the residuals vector, with $\M{s}$ as
given by \Eq{azek}.
To obtain the other partial derivatives we use the following relation.
For any parameter $\alpha$ we have
\begin{equation}
\pdrv{{\chi^2}}{\alpha}=\T {\left( \pdrv{{\chi^2}}{\M{F}} \right)}
\cdot \pdrv{\M{F}}{\alpha} \, .
\end{equation}
The first term $ \partial \chi^2 / \partial \M{F}$ is given by \Eq{drs} while the
second term reads, considering each unknown: 
\begin{eqnarray}
\pdrv{\M{F}}{\M{x}}&=& \diag(\Fext) \cdot \M{A} \, ,  \\
\pdrv{\M{F}}{\M{Z}}&=&\diag(\M{x}) \cdot \pdrv{\M{B}}{\M{Z}} \cdot
\diag(\Fext) \, , \\
\pdrv{\M{F}}{{E}}&=& \diag( \pdrv{\Fext}{E}) \cdot \M{A} \cdot \M{x} \, ,
\end{eqnarray}
with the same notation as in the appendix of the STECMAP paper.


\subsection{Gradients for the age-dependent kinematics recovery}
\label{s:aadk}
Again, we focus on the partial derivatives of the $\chi^2$.
Using \Eq{conf}, the model can be rewritten using the Fourier operator
\begin{equation}
\M{s}=\sum_{j=1}^{n} \for \cdot \diag(\fo \cdot \M{B}_j)
\cdot \fo \cdot \M{v}_j \, ,
\end{equation}
where $\M{B_{j}}$ is the discretized time-averaged {\SSP} of age $[t_{j-1},t_{j}]$,
The derivatives of $\chi^2$ relatively to $\M{v}$ can be derived directly from
\Eq{drsk} since the model is just a sum of convolutions. Replacing $\M{F} \leftarrow \M{B}_{j} $
and $\M{g} \leftarrow \M{v}_{j}$ yields the gradient of the $\chi^2$:
\begin{equation}
\pdrv{\chi^2}{\M{v}_{j}}=-2 \for \cdot \diag(\fo \cdot \M{F})^\ast \cdot \fo
\cdot \M{W
\cdot r} \, ,
\end{equation}
with the residuals vector $\M{r=y-s}$.
Finally, the derivative of $Q_{\mu}$ relatively to $\M{v}$ is the matrix defined by
\begin{equation}
\pdrv{Q_{\mu}}{\M{v}}=\left(
\pdrv{Q_\mu}{\M{v}_{1}} \,, \pdrv{Q_\mu}{\M{v}_{2}} \,
\dots \, \pdrv{Q_\mu}{\M{v}_{n}} 
\right) \, .
\end{equation}

\label{lastpage}

\end{document}